\documentclass{aa}  

\usepackage{natbib}
\bibpunct{(}{)}{;}{a}{}{,} 
\usepackage{graphicx}
\usepackage{color}
\usepackage{txfonts}
\begin{document}

   \title{Chemical abundance analysis of extremely metal-poor stars in the Sextans dwarf spheroidal galaxy
 \thanks{Study based on data collected with the Subaru Telescope, operated by the National Astronomical Observatory of Japan.}}


   \author{M. Aoki\inst{1}\fnmsep\inst{2},
          W. Aoki\inst{3},
          P. Fran\c{c}ois\inst{4}\fnmsep\inst{5}}

   \institute{European Southern Observatory, Karl-Schwarzschild-Str. 2, 85748 Garching bei Muenchen, Germany \\
              \email{maoki@eso.org}
         \and
               Ludwig-Maximilians-Universit{\"a}t M{\"u}nchen, Germany
         \and
             National Observatory of Japan, Mitaka, Tokyo, Japan\\
             \email{aoki.wako@nao.ac.jp}
         \and
             GEPI, Observatoire de Paris, PSL Research University, CNRS, Univ. Paris Diderot, Sorbonne Paris Cit{\'e}, 61 Avenue de l’Observatoire, 75014 Paris, France\\
             \email{patrick.francois@obspm.fr}
             \and
             Universit{\'e} de Picardie Jules Verne, 33 rue St Leu, Amiens, France 
             }

   \date{Received xx August, 2019; accepted xx March, 2020}

 
  \abstract
  { Metal-poor components of dwarf galaxies around the Milky Way could be remnants of the building blocks of the Galactic halo structure. Low-mass stars that are currently observed as metal-poor stars are expected to have formed in chemically homogeneous clusters in the early phases of galaxy formation. They should have already disintegrated and should exhibit large scatter in abundance ratios of some sets of elements (e.g., Sr/Ba) in the Milky Way field stars. However, chemical abundance ratios are expected to cluster in very metal-poor stars in dwarf galaxies because the number of clusters formed in individual galaxies in the very early phase is expected to be quite limited.} 
   {We examine the possible clustering of abundance ratios of Sr and Ba in the Sextans dwarf galaxy to test for the clustering star formation scenario.}
   {We investigate a total of 11 elements (C, Mg, Ca, Sc, Ti, Cr, Mn, Ni, Zn, Sr, Ba) in five stars in the Sextans dwarf galaxy. Previous studies suggest that these have similar abundance ratios. In this study, we focus on the abundance ratio of Sr to Ba. The observations are based on high-resolution spectroscopy (R = 40 000) using the Subaru Telescope High Dispersion Spectrograph. }
   {The distribution of $\alpha$/Fe abundance ratios of the Sextans dwarf galaxy stars is slightly lower than the average of the values of stars in the Galactic halo. The Sr/Ba abundance ratios for the five metal-poor stars are in good agreement, and this clumping is distinctive compared to the [Sr/Ba] spread seen in the metal-poor halo stars. We find that the probability of such clumping is very small if the Sextans stars have distributions of Sr and Ba abundances similar to halo stars.}
  {}

  \authorrunning{M.Aoki et al.}
  \maketitle
%
\section{Introduction}
According to the scenarios of structure formation, small galaxies like dwarf spheroidal galaxies have contributed to building up the larger ones, including the Milky Way \citep[e.g.,][]{diemand07}. Numerical studies such as that by \citet{font06} suggest that the accreted substructures should be detectable kinematically and chemically, even billions of years after the Milky Way first formed. Indeed, evidence in favor of this scenario is found in the difference in stellar dynamics, showing that the
halo is separated into substructures \citep[e.g.,][]{helmi99, stark09, xue11}. 
  \begin{table*}
      \caption[]{Object data}
         \label{Objectdata}
\centering
     $$
         \begin{tabular}{lccccccr}
            \hline
            \noalign{\smallskip}
               Star & RA & DEC &Exp.Time & S/N & S/N& Date&Ref \\
               &(J2000)&(J2000)&(s)&(4100Å)&(5180Å)&(UT)&\\
            \hline
            \noalign{\smallskip}
     S~10-14 & 10:13:34.70 & $-$02:07:57.9 & 14,200&14&56&2016 April 26&(1)\\
     S~11-13 & 10:11:42.96 & $-$02:03:50.4 &14,400 &21&74&2016 April 27&(1)\\
     S~49 &10:13:11.55& $-$01:43:01.8 &14,400&14&62 &2016 April 28&(2)\\        
                        \hline
         \end{tabular}
     $$
\tablebib{(1)~\citet{waoki09}; (2) \citet{shetrone01}
}
   \end{table*}

Another useful technique is the so-called chemical tagging, which aims to
assign stars to groups based on their chemistry \citep[e.g.,][]{freeman02}. Low-mass stars that are currently observed as metal-poor stars are expected to have formed in chemically homogeneous clusters in the early phases of the galaxy formation. The classification of
stars into groups with similar chemical composition is used to identify stars with a common origin, possibly in
the same cluster.
Nevertheless, very metal-poor stars ([Fe/H]$< -$2.5~dex) in the halo field exhibit a smooth dispersion in abundance ratios suggesting that a large number of such clusters have contributed to forming the halo structure. However, in order to apply chemical tagging to the Milky Way halo, a considerably large sample is required. There is an intention to
apply this technique to the stars of field halos by large-scale spectroscopic follow-up of the Gaia sample \citep[e.g.,][]{hawkinswyse18}. 

On the other hand, the application of chemical tagging to the metal-poor range of faint dwarf galaxies is expected to be
more straightforward. Faint dwarf galaxies contain very metal-poor stars whose chemical abundances are useful for studying
the environment and the formation process of galaxies. \citet{bland10} performed a detailed investigation of the formation of clusters with homogeneous chemical composition, and found that a small number of very metal-poor stars do not form smooth distributions but make clumps in the abundance plane of [Fe/H] versus [X/Fe], including the neutron-capture elements (e.g., [Ba/Fe]). This is in clear contrast to field halo stars, which would have also been born in clusters, but the number of
clusters is so large that their abundances are expected to become dispersed, leading to a large and smooth distribution in abundance ratios of elements. The clustering in elemental abundances of metal-poor
stars in dwarf galaxies is expected to be useful for examination of the procedure of chemical tagging and would help to constrain the formation scenario for the Milky Way.

\begin{sidewaystable*}
\caption{\label{tab:stepa} Stellar parameter and comparison with previous studies}
\centering
\begin{tabular}{lccccccccccl}
\hline\hline
            \noalign{\smallskip}
Star & $T_{\rm eff}$ \tablefootmark{1} & [Fe/H] & log {\it g}\tablefootmark{2} & $\xi$ & $\Delta$ $T_{\rm eff}$ &$\Delta$ [Fe/H] &  $\Delta \mbox{log} g$ & $\Delta$ $\xi$&V&K&Prev. Study\\
& (K) & (dex) & (dex) & (km $s^{-1}$)&(K)&(dex)&(dex)&(km s$^{-1}$)\\
\hline
            \noalign{\smallskip}
       S~10-14  &4620 & $-$2.82 & 1.02 &  2.52& 0  & $-0.12$&$-0.18$ & $+0.30$&17.64&15.08&\citet{waoki09}\\
    S~11-13 &4430 & $-$2.82 & 0.86 &  2.28 & $+30$  & $-0.02$& +0.26&$-0.12$&17.53&14.71&\citet{waoki09}\\
    S~49&4390 & $-$3.06 & 0.86 &  2.56& $+65$ & $-0.21$&$+0.76$  & $+0.06$&17.52&14.67& \citet{shetrone01}\\
\hline
            \noalign{\smallskip}
       S~24-72  &4340 & $-$2.90 & 0.74 &  2.72& $-90$  & $+0.03$&$-0.01$ & $+0.52$&17.35&14.42&\citet{tafel10}\\
    S~11-04 &4230 & $-$2.85 & 0.62 &  2.85 & $-90$& $+0.09$ &$+0.05$& $+0.65$&17.23&14.13&\citet{tafel10}\\
\hline
            \noalign{\smallskip}
    HD~88609 &4550& $-$2.97 & 0.91 &  2.60 & $0$& $+0.09$ &$-0.19$& $+0.08$&8.62&6.01&\citet{honda07}\\
\hline
\end{tabular}
\tablefoot{The difference is taken as our results minus other works.\\
\tablefoottext{1}{ from V-K \citep{gonz09}}\\
\tablefoottext{2}{computed from standard relation between absolute bolometric magnitude, temperature, and mass.}\\
}
\end{sidewaystable*}

The Sextans dwarf spheroidal galaxy would be an ideal galaxy for examination of chemical tagging. \citet{waoki09} show measurements of six metal-poor stars of the Sextans dwarf galaxy with low magnesium (Mg), calcium (Ca), and barium (Ba) abundance ratios. \citet{karlsson12} suggested the possibility that clustering with homogeneous chemical composition is apparent in this dwarf galaxy from the observation of Sextans metal-poor stars showing a clump in the [Mg/Fe] and [Fe/H] plane around [Fe/H]$\sim -$2.8. However, the
number of elements studied so far for chemical tagging is still relatively small. Further abundance measurement for metal-poor stars in the Sextans dwarf galaxy would be an ideal way to examine the usefulness of the chemical tagging method.

Chemical tagging is usually applied using abundance ratios of
$\alpha$-elements and Fe-peak elements because $\alpha$/Fe reflects the timescale of the chemical evolution of the system \citep[e.g.,][]{tinsley79}. However, the abundance differences between stars are not very large (at most 0.5~dex). Abundance ratios of neutron-capture elements (e.g., [Sr/Ba]) confer an advantage for chemical tagging because they show large scatter in their abundance ratios, and the differences can be clearly measured. For the Sextans dwarf galaxy,
chemical tagging using the abundance ratios of neutron-capture elements appears to be possible according to previous observations. There is a total of nine very metal-poor stars ($-3.0 <$ [Fe/H] $< -$2.6) for which Ba abundance has been measured in previous studies (two stars by \citet{tafel10}, six by \citet{waoki09}, and one by \citet{shetrone01}). Seven out of these very metal-poor stars show very good agreement of [Ba/Fe]$\sim -$1.2~dex. This clumping is remarkable, given the large scatter of [Ba/Fe] seen in the field halo stars in the same metallicity range. The two remaining stars, S~15-19 and S~12-28 \citep{waoki09}, have an excess of Ba. Furthermore, S~15-19 ([Ba/Fe] = 0.5~dex) is considered to be an s-process enhanced star \citep{honda11}. The similarity of the [Ba/Fe] in the remaining stars could be a signature of low-mass star formation in the same cluster, their Ba sharing the same origin.
\begin{table*}[h]
\begin{center}
\caption{Abundance changes from changing stellar parameters for S~49}
\begin{tabular}{lrrp{1pt}rrp{1pt}rrp{1pt}rr}
\hline
\hline
            \noalign{\smallskip}
& \multicolumn{2}{c}{$\Delta$ $T_{\rm eff}$ } & & \multicolumn{2}{c}{$\Delta$ [Fe/H]} & & \multicolumn{2}{c}{$\Delta $log  {\it g}} & & \multicolumn{2}{c}{$\Delta$ $\xi$} \\
\cline{2-3} \cline{5-6} \cline{8-9} \cline{11-12}
            \noalign{\smallskip}
   Species  & $+150$K & $-150K$ & & $+0.3$ & $-0.3$ & & $+0.3$ & $-0.3$ & & $+0.5$ & $-0.5$ \\
\hline
            \noalign{\smallskip}
    Mg I&$0.15$ &$-0.15$ & &$-0.02$ & 0.03 & &$-0.06$&$0.06$& &$-0.04$ & $0.06$ \\
    Ca I& $0.09$ & $-0.09$ & & $-0.01$ & 0.02& & $-0.03$ & 0.03 & & $-0.03$ & $0.06$ \\
    Sc II& $0.11$ & $-0.10$ & & $-0.03$ & 0.02& & 0.07 & $-0.06$ & & $-0.19$ & $0.21$ \\
    Ti I&$0.18$ & $-0.21$ & & $-0.01$ & $0.00$ & & $-0.04$ & 0.04 & & $-0.02$ & $0.04$ \\
    Ti II & $0.06$ & $-0.08$ & & $0.01$ & $0.00$ & & 0.07 & $-0.06$ & & $-0.13$ & 0.19 \\
    Cr I & $0.17$ & $-0.19$ & & $-0.05$ & $0.03$ & & $-0.09$ & $0.10$ & & $-0.05$ & $0.07$ \\
    Mn I & $0.18$ & $-0.19$ & & $-0.18$ & $0.16$ & & $-0.07$ & $0.08$ & & $-0.07$ & $0.07$ \\
    Fe I& $0.20$ & $-0.27$ & & $-0.04$ & $0.02$ & & $-0.05$ & 0.06 & & $-0.13$ & $0.19$ \\
    Fe II& $0.00$ & $-0.01$ & & $0.01$ & 0.00 & & $0.08$ & $-0.07$ & & $-0.12$ & $0.18$ \\
    Ni I& $0.14$ & $-0.14$& & $0.00$ & 0.04 & & $-0.04$ & 0.06 & & $-0.05$& $0.09$ \\
    Zn I& $0.04$ &$-0.03$& & $0.01$ & $-0.01$ & & 0.06 & $-0.04$ & & $-0.01$ & 0.03 \\
    Sr II& $0.13$ & $-0.12$ & & $-0.09$ & 0.07 & & $0.03$ & $-0.04$ & & $-0.20$ & 0.24 \\
    Ba II& 0.13 & $-0.10$ & & $0.00$ & 0.01 & & $0.04$ & $-0.02$ & & $-0.06$& 0.09 \\
\hline 
 \end{tabular}
\tablefoot{The difference is taken as the abundance measured after changing the stellar parameters minus our final abundance. For Mn, the abundance difference from S~11-13 are taken.\\
}
\label{tab:error}
\end{center}
\end{table*}

Moreover, the Sr abundance of two of these stars was measured by \citet{tafel10} using the Very Large Telescope (VLT). The abundance ratios [Sr/Ba] of the two stars are in very good agreement, measuring  0.89~dex and 0.84~dex for S~24-72 and S~11-04, respectively. The Milky Way halo stars show a large and smooth dispersion of [Sr/Ba] ($\geq$ 2~dex) for field halo stars of the same metallicity and in a similar [Ba/Fe] range.  We therefore expect that determination of the abundance of Sr and subsequent determination of the [Sr/Ba] ratio provides the strongest constraint on the model of chemical clustering in dwarf galaxies.

In \S2, we describe the sample selection and the details of spectroscopic observations. \S3 gives the estimates of the stellar parameters and the details of the chemical abundance analysis. In \S4, we present our results. We discuss the derived abundances in \S5. Finally, we summarize our study in \S6. 
\section{Observation}
Metal-poor stars in the Sextans dwarf spheroidal galaxy were selected for our study to obtain high-resolution spectra of the UV-blue range. We selected stars that have similar Ba abundances according to previous studies by \citet{waoki09} and \citet{shetrone01}. The selected stars have similar metallicity to the two stars for which the Sr abundance was measured by \citet{tafel10} ([Fe/H]$\sim-2.8$). We selected S~10-14 and S~11-13 from \citet{waoki09} and S~49 from \citet{shetrone01}, as they are the three brightest stars (V$\sim17.5$) among the target candidates.

The targets were observed from 2016 April 26 to 28 for the first half of the night for all three days with the 8.2 m Subaru Telescope High Dispersion Spectrograph (HDS, \citet{noguchi02}). The wavelength coverage is from 3920 to 5604~\AA~with a resolving power of  {\it R}=40,000 (0.9 arcsec slit). The signal-to-noise ratio (S/N) per resolution element (3.7 pixels) of the spectrum is estimated from photon counts at 4100~\AA~ and 5180~\AA. Positions of objects, exposure time, S/N, and observed dates are summarized in Table~\ref{Objectdata}. 

We reduced the raw data via a standard process using the IRAF {\'e}chelle package\footnote{IRAF is distributed by National Optical Astronomy Observatories, which are operated by the Association of Universities for Research in Astronomy, Inc., with the cooperation of the National Science Foundation}. The effect of the sky background is significant in spectra that were taken at the end of the observation when the moon rose. We removed the sky background from the spectra by extracting them from the region around the stellar spectra on the slit. The individual spectra were then combined after the wavelength calibration. 

\section{Chemical abundance analysis}
\begin{sidewaystable*}
\begin{center}
\caption{Elemental abundances}
\footnotesize
\begin{tabular}{llcccccccccccccccc}
\hline
\hline
    \noalign{\smallskip}
Star& Elem. &FeI&FeII&C&MgI&CaI&ScII&TiI&TiII&CrI&MnI&NiI&ZnI&SrII&YII&BaII&EuII\\
\hline
            \noalign{\smallskip}
S~10-14&log$\epsilon$&4.68&4.38&5.69&4.85&3.65&0.05&$<$2.27&2.10&2.43&2.71&3.40&. . .&$-1.00$&$< -$ 0.96&$-2.16$&. . .\\
&N&29&3&2&3&1&1&3&5&3&2&1&. . .&2&1&2&. . .\\
&[X/Fe]&. . .&. . .&0.08&$0.07$&$0.13$&$-$0.28&$<$0.14&$-0.03$&$-0.39$&0.10&$0.00$&. . .&$-1.05$&$< -$0.35&$-1.52$&. . .\\
&$\sigma /\sqrt(N)$&0.06&0.05&. . .&0.10&0.31&0.31&&0.12&0.17&0.22&0.31&. . .&0.22&&0.22&. . .\\
&err&. . .&. . .&0.32&0.21&0.39&0.35&&0.24&0.23&0.30&0.36&. . .&0.28&&0.30&. . .\\
\hline
            \noalign{\smallskip}
S~11-13&log$\epsilon$&4.68&4.84&5.34&4.93&3.70&0.12&2.16&2.17&2.30&2.15&3.34&2.25&$-1.34$&$< -$1.24&$-2.30$&. . .\\
&N&55&4&2&3&3&4&3&14&4&1&1&1&1&1&2&. . .\\
&[X/Fe]&. . .&. . .&$-0.27$&$0.15$&0.18&$-$0.21&$0.03$&$0.04$&$-0.52$&$-0.46$&$-0.06$&0.51&$-1.39$&$< -$0.63&$-1.66$&. . .\\
&$\sigma /\sqrt(N)$&0.04&0.09&. . .&0.11&0.05&0.04&0.05&0.05&0.07&0.29&0.29&0.29&0.29&&0.21&. . .\\
&err&. . .&. . .&0.31&0.21&0.24&0.17&0.18&0.20&0.17&0.36&0.34&0.41&0.34&&0.29
&. . .\\
\hline
            \noalign{\smallskip}
S 49& log $\epsilon$&4.44&4.56&5.15&4.74&3.57&$-$0.02&$<$1.96&2.03&2.14&. . .&3.05&2.02&$-1.08$&. . .&$-2.13$&$<-$2.25\\
&N&29&3&2&3&3&2&3&7&3&. . .&1&1&2&. . .&1&1\\
&[X/Fe]&. . .&. . .&$-0.22$&0.20&0.29&$-$0.11&$<0.07$&0.14&$-0.44$&. . .&$-0.11$&0.52&$-0.89$&. . .&$-1.25$&$< $ 0.29\\
&$\sigma /\sqrt(N)$&0.04&0.08&. . .&0.12&0.03&0.17&&0.08&0.10&. . .&0.24&0.24&0.17&. . .&0.24&\\
&err&. . .&. . .&0.31&0.22&0.23&0.24&&0.21&0.18&. . .&0.30&0.38&0.24&. . .&0.32&\\
\hline
\hline
            \noalign{\smallskip}
S 24-72& log $\epsilon$&4.60&4.45&6.40&4.88&3.54&0.11&1.85&2.28&2.41&2.22&3.35&2.09&$-1.03$&. . .&$-2.04$&$<-$2.60\\
&N&45&3&2&3&4&2&4&5&3&3&1&1&1&. . .&3&1\\
&[X/Fe]&. . .&. . .&0.87&0.18&0.10&$-$0.14&$-0.20$&0.23&$-0.33$&$-0.31$&0.03&0.43&$-1.00$&. . .&$-1.32$&$<-$0.22\\
&$\sigma /\sqrt(N)$&0.03&0.09&. . .&0.05&0.07&0.14&0.07&0.04&0.06&0.05&0.20&0.20&0.20&. . .&0.12&\\
&err&. . .&. . .&0.36&0.19&0.24&0.22&0.18&0.20&0.16&0.21&0.26&0.35&0.26&. . .&0.23&\\
\hline
            \noalign{\smallskip}
S 11-04& log $\epsilon$&4.66&5.17&4.84&5.03&3.68&0.17&2.04&2.38&2.56&2.09&3.12&2.09&$-0.62$&. . .&$-1.67$&$<-$2.67\\
&N&44&3&2&3&5&2&8&5&4&2&1&1&2&. . .&2&1\\
&[X/Fe]&. . .&. . .&$-0.74$&0.28&0.19&$-0.13$&$-0.06$&$0.28$&$-0.23$&$-0.49$&$-0.25$&0.38&$-0.64$&. . .&$-1.00$&$<-$0.34\\
&$\sigma /\sqrt(N)$&0.02&0.01&. . .&0.025&0.05&0.11&0.04&0.04&0.01&0.11&0.15&0.15&0.11&. . .&0.11&\\
&err&. . .&. . .&0.33&0.18&0.23&0.19&0.17&0.20&0.15&0.23&0.23&0.33&0.20&. . .&0.23&\\
\hline
\label{tab:abund}
\end{tabular}
\end{center}
\end{sidewaystable*}

Chemical abundances are determined based on model atmospheres and spectral line data. We employ the ATLAS model atmospheres with the revised opacity distribution function (NEWODF) by \citet{caskur03}. 
We applied the one-dimensional local thermodynamic equilibrium (LTE)
spectral synthesis code, which is based on the same assumptions as the
model atmosphere program of \citet{tsuji78} and has been used in
previous studies \citep[e.g.,][]{waoki09b}.
The line list is given in Table~\ref{tab:linelist}. 

\subsection{Stellar parameters}
Among the stellar parameters, we estimate effective temperature ($T_{\rm eff}$) from the color ($V-K$), adopting the $K$ magnitude and $V$ magnitude from the SIMBAD astronomical database\footnote{SIMBAD Astronomical Database: http://simbad.u-strasbg.fr/simbad/ } \citep{wenger00} for the three target stars. We used $V-K$ since the temperature scales are less dependent on metallicity and molecular absorption in giant stars. We estimated $T_{\rm eff}$ from the color--temperature relation for giant stars by \citet{gonz09}. Different extinction for foreground reddening was estimated for different stars in the range 0.01$<$E(B$-$V)$<$ 0.05. The uncertainty of  $T_{\rm eff}$ due to photometry errors (about 0.1~mag) and uncertainty of reddening (0.05~mag) is about 100~K. Including the uncertainty of the  $T_{\rm eff}$ scale, we adopt 150~K as the uncertainty that is applied to estimate abundance errors.

 Surface gravity (log {\it g}) was determined using the following relation with effective temperature, mass, and bolometric magnitude:
    \begin{equation}
\log{\it g}_{\mathrm{*}}=\log{\it g}_{\odot}+\log \frac{M_{\mathrm{*}}}{M_\odot}+4\log \frac{T_{\mathrm{eff *}}}{T_{\mathrm{eff}\odot}}+0.4 \left( M_{\mathrm{Bol*}}-M_{\mathrm{Bol\odot}}\right),
   \end{equation}
 where log {\it g}$_{\odot}=4.44$,  $T_{\rm eff \odot}=5790$K, and $M_{\rm Bol \odot}=$4.74 for solar values and $M_{\rm *}=0.8{M_\odot}$ for the mass of the RGB stars are adopted. We calculated the absolute bolometric magnitude ($M_{\rm Bol *}$) of the stars using the calibration for the bolometric correction
from \citet{alonso99}. We assume 90 kpc for the distance to the Sextans dwarf galaxy \citep{karach04}. We estimated the uncertainty of log {\it g}  adopting errors of stellar mass (10$\%$),  $T_{\rm eff}$ (150~K), and $M_{\rm bol}$ (0.3~mag), resulting in 0.22~dex. We adopt 0.3~dex as the uncertainty that is applied to estimate the abundance error.
 
We adjusted micro-turbulence ($\xi$) so that the Fe abundances derived from individual lines do not show systematic differences depending on the strength of the \ion{Fe}{I} lines. By changing $\xi$, the trend appears in Fe abundance against the equivalent width. The uncertainty of $\xi$ is estimated when this trend in the Fe abundance becomes larger by 1 $\sigma$.

Finally, we determined metallicity ([Fe/H]) from the final averaged abundance of \ion{Fe}{I}. The errors are estimated from the scatter of Fe abundances derived from individual Fe I lines. Stellar parameters of the targets and their comparison with previous studies are summarized in Table~\ref{tab:stepa}. Some of the stellar parameters are different from those found in previous studies, which would affect the results of chemical abundances. The difference in chemical abundance and possible effects of the stellar parameters are discussed in the sections below. 

We re-analyzed the spectrum of HD~88609 obtained by \citet{honda07}.
  This is a cool red giant with a similar metallicity to the stars of our Sextans
  sample, and was well studied by previous works. The effective temperature for this object is taken from \citet{honda07},
  and other parameters are derived from the analysis of the
  high-resolution spectrum as done for the Sextans sample. The stellar parameters of this star obtained here are compared to those obtained by \citet{honda07} in Table \ref{tab:stepa}.

\subsection{Abundance measurements and error estimates}
In our error estimates, we investigate the systematic difference in chemical abundances that occurs with uncertainties of model atmosphere parameters. The effects of changes of stellar parameters on the abundances are given in Table~\ref{tab:error} for S~49. We expect that the other stars from the sample show similar behavior because they have similar stellar parameters, and we apply these parameters for error estimates of all the stars of our sample. Table~\ref{tab:error} shows differences in abundance measurements by changing $\pm$150 K for $T_{\rm eff}$, $\pm$0.3~dex for log {\it g}, $\pm$0.3~dex for [Fe/H], and $\pm$0.5 km$^{-1}$ for $\xi$. The abundance of elements increases when a higher $T_{\rm eff}$ is assumed, while it decreases when a higher $\xi$ is assumed. On the other hand, the effects of [Fe/H] and log {\it g} are generally different for abundances of neutral and ionized species. The abundances derived from neutral species are generally lower when higher [Fe/H] and log {\it g} are adopted, while the opposite is true for the ionized species. The exceptions to this rule are the abundances derived from \ion{Zn}{I} and \ion{Sr}{II} for changes of [Fe/H]. 

We measured the abundances of elements from magnesium (Mg) to barium (Ba) in the present study. Details of the abundance measurements for individual species are given in Sects. 3.3--3.5. The abundance of Sr is measured for the first time in all three targets. For those elements with several absorption lines available in our spectra, the averaged abundances from individual lines are taken as the final results. For those that have only one or a few measurable lines, we applied the spectrum synthesis technique. The effects of hyperfine splitting were taken into account in the analysis of Ba assuming the r-process isotope ratios \citep{mcwill98}. To derive the [X/Fe] values, we used the Solar-System abundances obtained by \citet{asp09}. Abundance and equivalent width of individual lines are given in Table~\ref{tab:linelist} and the final abundances adopted are summarized in Table~\ref{tab:abund}. In the following sections, we compare our abundances to those found in previous studies \citep[]{waoki09, shetrone01}.

The errors of derived abundances are estimated from the random
  errors and those due to uncertainties of stellar parameters. Random errors estimated for each species are $\sigma /\sqrt(N)$, where $\sigma$ is the standard deviation of abundances derived from individual lines and $N$ is the number of lines used (Table \ref{tab:abund}). For elements that have only one or two available lines, the standard deviation of Fe from individual Fe I lines is adopted.  The error due to uncertainties of stellar parameters is estimated for [X/Fe] values for element X. Namely, the changes of [X/Fe] by changing stellar parameters are calculated using the results given in Table 3. We also derive the error of [Sr/Ba] in the same manner. The random errors and errors due to uncertainties of stellar parameters are added in quadrature to derive the total errors in our analysis.

To verify the consistency of our analysis, we apply the adopted stellar abundance analysis technique to the thoroughly investigated star, HD~88609. There is no significant difference between our measurement and those of previous studies of HD~88609 \citep{honda07}. 

 For C abundances, we estimate the errors by spectrum synthesis of the CH bands. We include the change of the abundance due to possible changes in continuum level, and the effect of changing the $T_{\rm eff}$ by $\pm$150 K.

\subsection{Re-analysis of S~24-72 and S~11-04}
\citet{tafel10} measured the chemical abundance including Sr and Ba in S~24-72 and S~11-04. These latter authors obtained their spectra with the high-dispersion spectrograph UVES at VLT, and the abundance ratios of [Sr/Fe] and [Ba/Fe] in the two stars show good agreement. To combine the available data for these two stars with our results for S~10-14, S~11-13, and S49, we apply our analysis procedure to the UVES spectra of S~24-72 and S~11-04 provided by the ESO archive. The high-resolution spectra used for this re-analysis were obtained from Program ID 079.B-0672A and 081.B-0620A.

We normalized the data in the same manner as for HDS data using IRAF, and applied the ATLAS/NEWODF models for abundance measurements. The stellar parameters were also estimated in the same manner; $T_{\rm eff}$ from the ($V-K$), adopting the $K$ magnitude and $V$ magnitude from \citet{tafel10}. Here, log {\it g} was calculated from the photometric relation and [Fe/H] and $\xi$ were derived from the standard LTE analysis of \ion{Fe}{I} and \ion{Fe}{II} lines. The $T_{\rm eff}$ of our estimation is 90~K lower than that of \citet{tafel10} for both stars. As we applied the same calculation to derive log {\it g} as \citet{tafel10}, the difference in log{\it g} is very small. We also measured the chemical abundance using the line list adopted for our three target stars instead of using the line list adopted by \citet{tafel10}. 

\citet{tafel10} used DAOSPEC\footnote{DAOSPEC was written by P. Stetson for the Dominion Astrophysical Observatory of the Herzberg Institute of Astrophysics, National Research Council, Canada.} to normalize their data, and used the MARCS\footnote{http://marcs.astro.uu.se/} spherical model atmosphere. \citet{tafel10} estimated the photometric temperature from $V-I$, $V-J$, $V-H,$ and $V-K$ using the calibration of \citet{rami05}. \citet{tafel10} also restricted the Fe lines to excitation potentials larger than 1.4 eV, because \ion{Fe}{I} lines with low excitation potential could be affected by NLTE effects. Our estimated stellar parameters and comparisons with \citet{tafel10} are summarized in Table~\ref{tab:stepa}. 

 In the following sections, we compare the abundances of individual elements with the abundances found by \citet{tafel10}, along with our three targets.
   \begin{figure}
   \centering
\includegraphics[width=8.5cm]{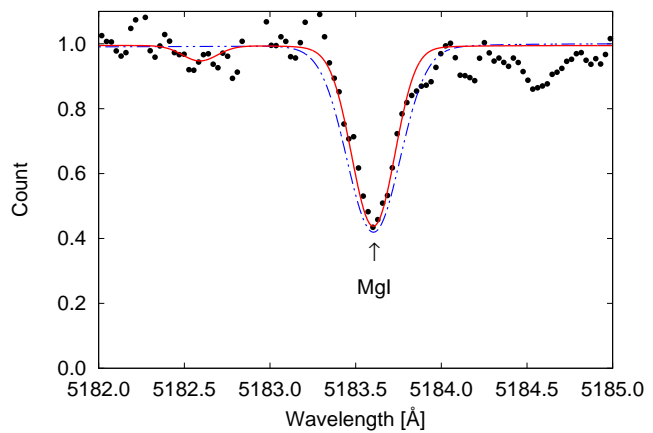}
      \caption{Observed Mg line of S~49 at 5183 Å (dots). The red solid line shows our synthetic spectra fitting. The blue dashed line shows the calculated line using the stellar parameter and the equivalent width derived by \citet{shetrone01}.}
         \label{fig:lines}
   \end{figure}
   \begin{figure*}
   \centering
\includegraphics[width=8.5cm]{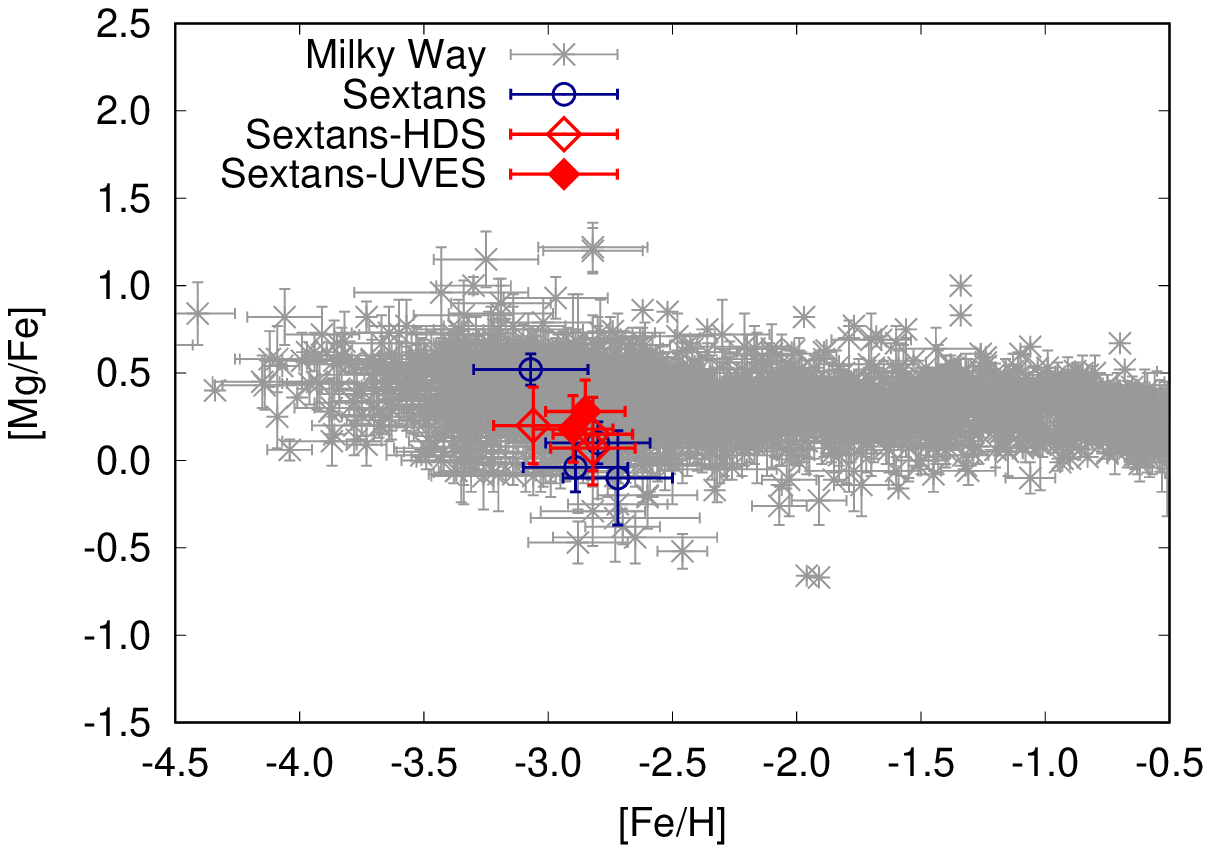}
\includegraphics[width=8.5cm]{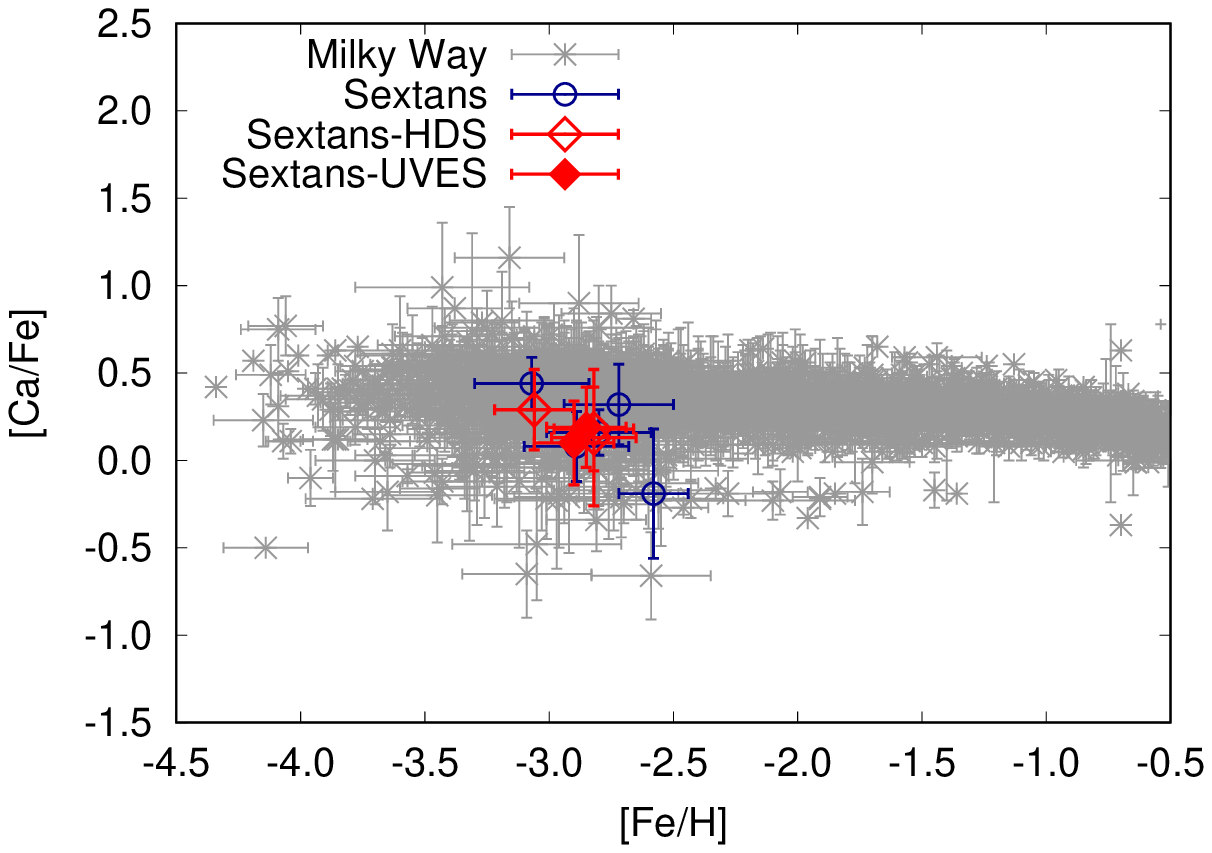}
      \caption{[Mg/Fe] and [Ca/Fe] as a function of [Fe/H]. The Sextans stars measured in this study are shown by diamonds with error bars, while previously studied Sextans stars ([Fe/H]$<-2.5$) are depicted as circles \citep{waoki09, honda11, kirby10}. The abundances of Galactic halo stars are shown by asterisks taken from the SAGA database \citep{suda08}.}
         \label{mgfe}
   \end{figure*}
\subsection{Elements up to neutron capture elements}
\subsubsection{Carbon}
Carbon abundance is measured using spectrum synthesis of the CH band at 4315~\AA~
and 4324~\AA. We estimate the [O/Fe]=0, and adopt the CH line list by \citet{masseron14}. The C abundance ratios of the three stars observed with Subaru are close to solar abundance ratio or lower  overall, and are comparable to those of evolved stars with lower temperature and lower gravity. The abundance ratio of S~11-04 is also low ([C/Fe]=$-0.74$), and is comparable to what \citet{tafel10} measured ([C/Fe]=$-0.91$). S~24-72 is the only C-enhanced star in our sample ([C/Fe]=0.87), and its enhancement is as also suggested by \citet{tafel10} ([C/Fe]=0.49).

\subsubsection{$\alpha$$-$elements}

   \begin{figure}
   \centering
\includegraphics[width=9cm]{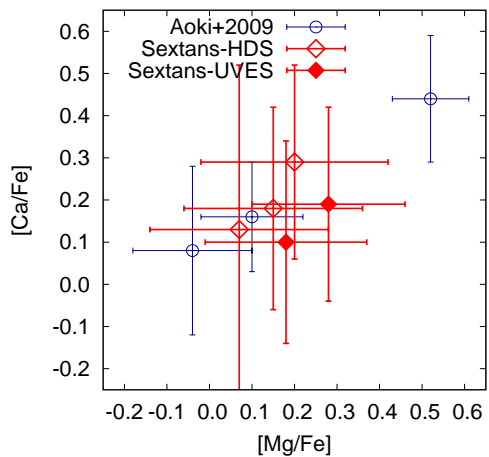}
      \caption{[Ca/Fe] as a function of [Mg/Fe]. The Sextans stars measured in this study are shown by diamonds with error bars, while measurements by \citet{waoki09} are depicted as circles. }
         \label{fig:mgvsca}
   \end{figure}
We measured Mg, Ca, and Sc as elements that represent the $\alpha$$-$elements. The Mg abundances were determined from one to three lines of 4571, 4702, 5183, and 5528 \AA. The line at 5172 \AA~is not used because of the effect of the bad column on the CCD. The abundances from individual lines show fairly good agreement with each other.
The line at 5183 \AA~ is the only one commonly used by the two latter-mentioned studies and ours. The equivalent widths of 5183 \AA~we measured for S~10-14 and S~11-13 are very similar, with a difference of around $\pm$5 m\AA~ from those obtained by \citet{waoki09}. Other lines commonly used for our measurement and that of \citet{waoki09} also have similar equivalent widths, with the largest difference being $\sim$15 m\AA~ for 4571 Å in S~11-13. The [Mg/Fe] ratios of S~10-14 and S~11-13 are $\sim 0.20$~dex larger than those reported by \citet{waoki09}, but we still confirm the low [Mg/Fe] ratios compared to the typical values for Galactic halo stars (e.g., $\sim0.4$~dex). On the other hand, \citet{shetrone01} report a high [Mg/Fe] = 0.41 for S~49 due to their large equivalent width, while we obtain a relatively low abundance ratio for this star as well ([Mg/Fe] = 0.20). This large difference is partially explained by the relatively large difference in the adopted stellar parameters. According to the error estimates (Table \ref{tab:error}), our abundance could be lower than that of  \citet{shetrone01} by $-$0.06~dex due to the adopted parameter. Another possible reason for the difference is the equivalent width that \citet{shetrone01} measured. 
The equivalent widths of our measurement for the 4702 \AA~and 5183 \AA~ lines are 48.5 m\AA~ and 186.5m\AA, while \citet{shetrone01} measure 88 m\AA~ and 229.5 m\AA, respectively. Figure~\ref{fig:lines} shows the observed Mg line of S~49 at 5183 \AA~compared with a synthetic spectrum corresponding to the equivalent width of ours and that of \citet{shetrone01}. We suspect that the measurement of the equivalent width of the line used by \citet{shetrone01} is an overestimate.

For S~24-72 and S~11-04, we used three lines (5172, 5183, and 5528 \AA) to derive the Mg abundances. \citet{tafel10} also used the same lines, and our abundance ratios [Mg/Fe] agree very well with their results. The [Mg/Fe] for the two stars also agree with the [Mg/Fe] of the three targets observed with Subaru HDS. The abundance ratios including other low-metallicity stars in the Sextans dwarf galaxy and field halo stars are shown in Figure~\ref{mgfe}.

The Ca abundances of the three stars observed with Subaru are derived from one to three Ca I lines (4454, 4455, 5265, 5588 \AA), while in S~24-72 and S~11-04, we use three and five lines, respectively, in $\lambda >$ 5588 \AA. The abundance ratios compared with other stars in the Sextans dwarf galaxy and halo stars are shown in Figure~\ref{mgfe}. There is no line in common with the previous studies since \citet{waoki09} and \citet{shetrone01} use lines in the longer wavelength range ($>$ 6000 \AA). 
 [Ca/Fe] of S~49 obtained by our analysis is higher than that of \citet{shetrone01} ([Ca/Fe]$=0.08\pm0.23$). However, this could be due to the lower Fe abundance derived by our study. The log$\epsilon$(Ca) is similar to their result. 

\citet{waoki09} discussed in their paper that the stars with low [Ca/Fe] ratios are found with low [Mg/Fe] values.  Our measurements of [Ca/Fe] and [Mg/Fe] reveal a similar trend, although our [Mg/Fe] and [Ca/Fe] for S~10-14 and S-11-13 are both slightly higher than those of \citet{waoki09}. Figure \ref{fig:mgvsca} shows this trend for the five stars that we measure in this work along with the other stars measured by \citet{waoki09}. The object with the highest Mg and Ca abundances is S15-19, which is well separated from the others \citep{waoki09} and has abundance ratios typically found in the Milky Way.

The Sc abundances are determined from one to four lines of 4314, 4374, 4400, and 4415~{\AA} for the HDS targets. The abundances from individual lines show relatively good agreement with each other. \citet{waoki09}  only measured the Sc abundance for S~11-13, using a line at 5526~{\AA}. This line is not measurable in our data. The [Sc/Fe] ratio of the star is $\sim 0.07$~dex smaller than that reported by \citet{waoki09}. \citet{shetrone01} do not report on the Sc abundance of S~49. Therefore, this study is the first to report the Sc abundance of S~10-14 and S~49. 

For S~24-72 and S~11-04, we used 5031 and 5526~{\AA} to derive the Sc abundances. \citet{tafel10} also used the same lines. The equivalent widths agree very well with the result of these latter authors. The difference in [Sc/Fe] ratio is around 0.06~dex, indicating that our results agree very well with theirs.
   \begin{figure}
   \centering
\includegraphics[width=8.5cm]{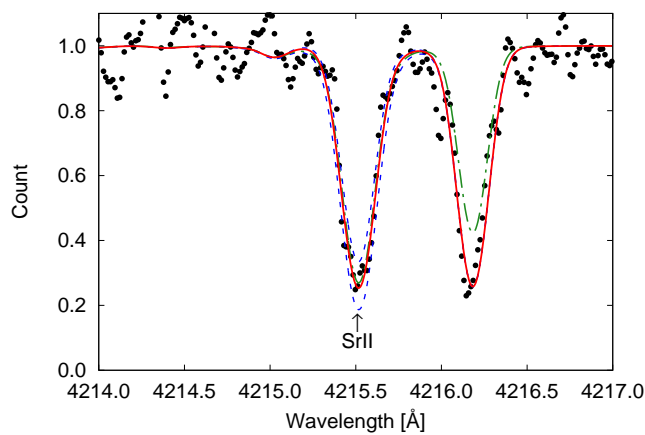}
      \caption{Observed Sr line of S~11-04 at 4215 Å (dots). The red solid line shows the synthetic spectra for the abundance derived by our analysis. The blue dashed lines are the spectra calculated after changing the Sr abundance by $\pm$0.5~dex. The green dot-dashed line shows a spectrum calculated as the equivalent width obtained by \citet{tafel10} is reproduced.}
         \label{fig:srline}
   \end{figure}

\subsubsection{Fe group elements}
The abundance of Ti is only obtained from species in the neutral and first ionization stages. Only the upper limit of Ti abundance is obtained for S~10-14 and S~49 from \ion{Ti}{I} lines. The abundance ratios from \ion{Ti}{I} and \ion{Ti}{II} lines of S~11-13 are in good agreement with each other. For S~49, the abundance ratio [Ti/Fe] by \citet{shetrone01} is $-0.29\pm0.15$~dex, which is notably lower than what we obtained, even considering the difference in adopted stellar parameters. No \ion{Ti}{I} lines were measured that are common to both our study and that of \citet{shetrone01}. 

In the study by \citet{tafel10}, large differences were seen in the first and second ionization stages for S~24-72 and S~11-04; these latter authors find that the ratio of [Ti/Fe] from \ion{Ti}{II} lines is $\sim0.42$~dex larger than that from \ion{Ti}{I} lines. Our measurements for S~24-72 and S~11-04 also show similar differences (Table \ref{tab:abund}). 

The \ion{Cr}{I} lines at 4254, 5345 and 5409~{\AA} are used for all three targets, whereas 4289 Å is also used for S~11-13. The abundance ratios of [Cr/Fe] of the three stars are in good agreement with each 
other. The Cr abundance of S~24-72 and S~11-04  was measured using three or four lines at 5256, 5208, 5345, and 5409~{\AA}. The Cr abundances of all five stars are in good agreement with those of previous studies, and we also note that the abundance ratios are in good agreement with those of metal-poor red giant stars in the Galactic halo.

The Mn abundance was measured using one or two lines at 4048, 5407, and 5420 \AA~ for S~10-14 and S~11-13. \citet{waoki09} do not report on Mn abundance. The Mn abundance of S~49 cannot be measured in our study, while \citet{shetrone01} measure its upper limit. The [Mn/Fe] ratio of S~10-14 is higher than that of the other targets including S~24-72 and S~11-04, but the ratio is still within the typical values for  metal-poor Galactic halo stars.

The Mn abundance  of S~24-72 and S~11-04  was measured using two to three lines at 4041, 4783, and 4823 \AA. The abundances measured using individual lines are in good agreement with each other. \citet{tafel10}  only used the line at 4823 \AA, and the equivalent widths measured by these latter authors agree well with our measurements. The [Mn/Fe] ratios are also in good agreement.

The abundance of Ni was measured using a single \ion{Ni}{I} line at 5476 Å for all three targets. This was also done for S~11-13 and S~49 by \citet{waoki09} and \citet{shetrone01}. The abundance ratio [Ni/Fe] that we obtain is lower for both stars compared to these two latter-mentioned studies. For S~49, the difference is larger than the error. The equivalent width that these latter two groups measured for 5476 Å is 116 mÅ, which is $\sim$46 mÅ larger than ours. As mentioned for Mg, the equivalent widths measured by  \citet{shetrone01} are overall larger than our measurements, and lead to larger abundances (except for Ca; but the difference is within the error and could be due to the difference in metallicity). 

Using the same single line, we measured the abundance of Ni  for S~24-72 and S~11-04.  Our [Ni/Fe] ratios are smaller than those of  \citet{tafel10} by $-$0.20 and $-$0.19 for S~24-72 and S~11-04, respectively. Our equivalent widths measured for both stars are in good agreement. The difference in [Ni/Fe] of S~24-72 can be explained by the difference in applied stellar parameters. A relatively large difference is found in $\xi$ and $T_{\rm eff}$ which results in a lower Ni abundance according to our error estimate (Table \ref{tab:error}). 

We obtained the abundance of Zn  for two targets, S~11-13 and S~49, with a single \ion{Zn}{I} line at 4810 Å. This element was not measured by \citet{waoki09} and only an upper limit was estimated for S~49 by \citet{shetrone01} using the same line.
The same line was used for S~24-72 and S~11-04 to obtain the Zn abundance in the present study, whereas \citet{tafel10} did not measure this element. Overall, the abundance ratios are in good agreement with those of metal-poor red giant stars in the Galactic halo.

\subsection{Neutron-capture elements}
We obtained Sr and Ba abundances for our targets in the Sextans dwarf galaxy and comparison stars.  We also estimated the upper limits of Y and Eu. The results are given in Table \ref{tab:abund}. The upper limits estimated for the two elements are not very meaningful because of the limited quality of our spectra and the low abundances expected for these elements from Sr and Ba abundances. An overall summary of Sr and Ba abundances is given in Table~\ref{tab:srba}.

We measured the abundance of Sr  using the two \ion{Sr}{II} lines at 4078 Å and 4216 Å, which is the first time this has been achieved for S10-14, S~11-13, and S~49. The log$\epsilon$(Sr) of the three stars are in good agreement with each other. We used the same two lines to obtain the Sr abundance of S~24-72 and S~11-04. The [Sr/Fe] of S~24-72 is in good agreement with that of three of the stars in the sample, while the [Sr/Fe] of S~11-04 is slightly higher than the other four stars. The [Sr/Fe] values measured by \citet{tafel10} for S~24-72 and S~11-04 are $-0.21\pm0.35$ and  $-0.01\pm0.40$, respectively, which are considerably larger than what we obtain for the same stars, even considering the difference in the stellar parameters. 

Figure~\ref{fig:srline} shows the Sr line for S~11-04 at 4215 Å compared with the synthetic spectra corresponding to the equivalent widths of ours and of \citet{tafel10}.  The equivalent width  that we measure for Sr  is not very different from that measured by these latter authors, suggesting
  that the difference in the stellar parameters could be the reason for the difference in abundance.

According to our error estimates (Table \ref{tab:error}), the abundance of \ion{Sr}{II} is relatively sensitive to stellar parameters. The difference in the value of the parameters between the two studies could result in a difference in Sr abundance for S~24-72 of $\sim -$0.34~dex according to our error estimates. Similarly, for S~11-04, the relatively large difference in its parameters could change the Sr abundance of S~11-04 by $-$0.40~dex. However, these differences cannot fully explain the large difference between the [Sr/Fe] that we obtain and that of \citet{tafel10}, and a 0.4~dex difference remains. 

   \begin{figure}
   \centering
\includegraphics[width=8.5cm]{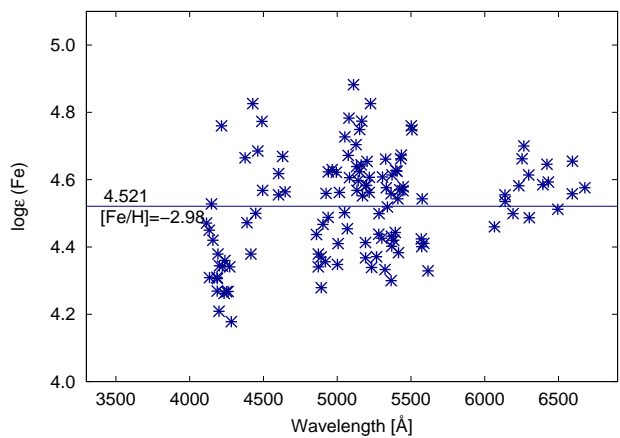}
      \caption{Abundance of \ion{Fe}{I} derived from individual lines in HD~88609 as a function of wavelength. There is no trend of abundance with wavelength, since scattering is taken into account in our model.}
         \label{fig:logewavel}
   \end{figure}
   \begin{figure*}
   \centering
\includegraphics[width=5.5cm, angle=270]{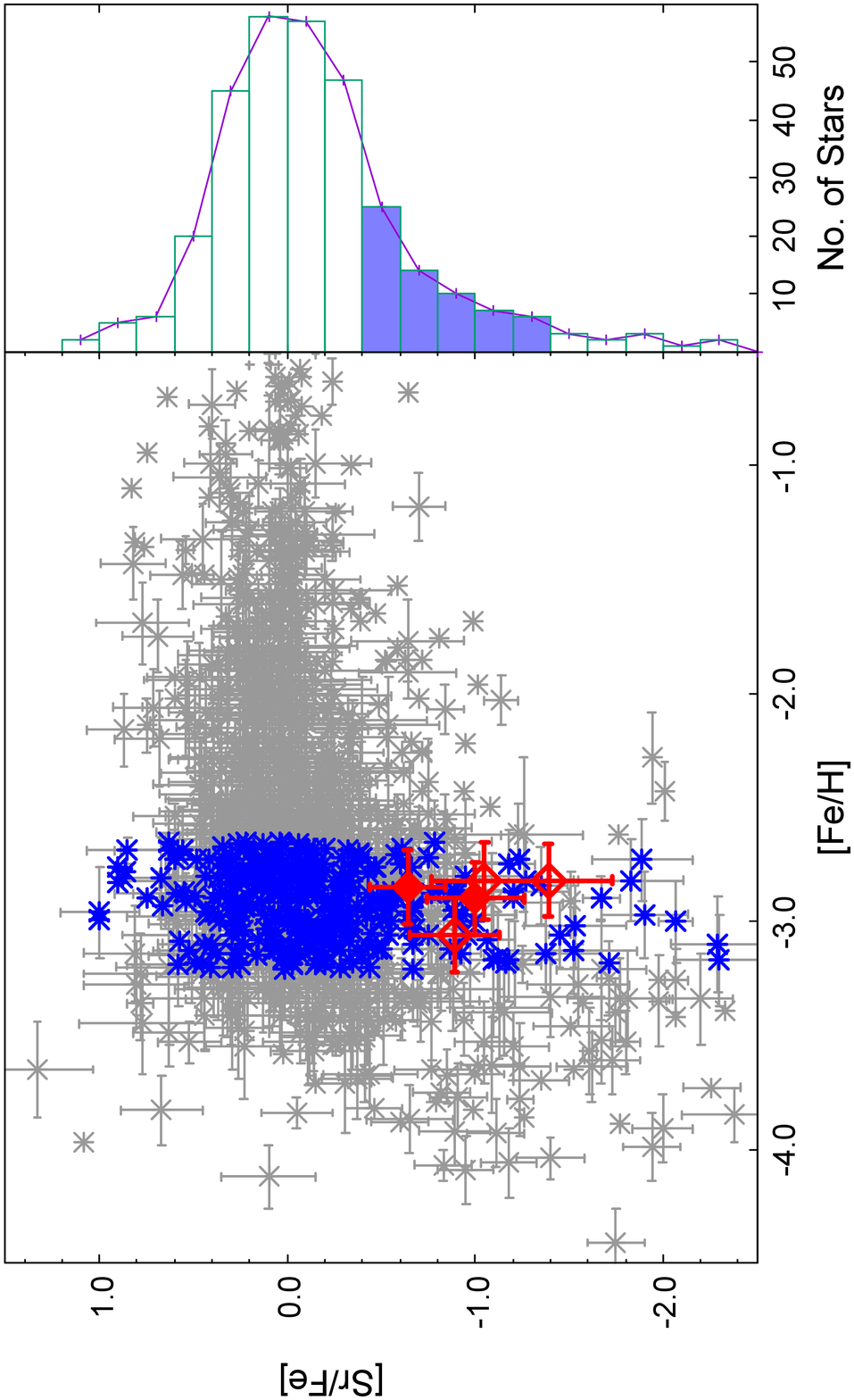}
\includegraphics[width=5.5cm, angle=270]{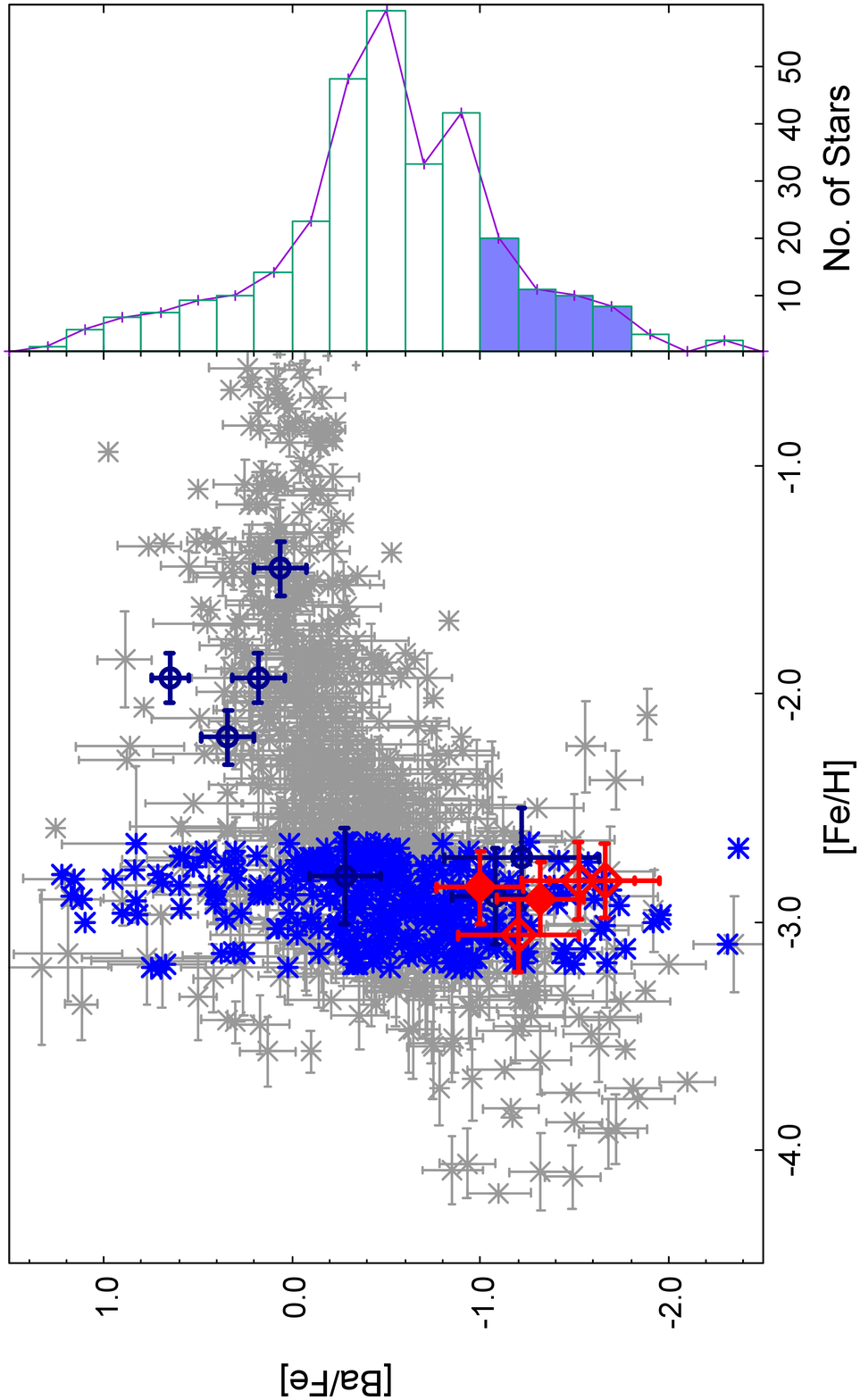}
      \caption{[Sr/Fe] and [Ba/Fe] as a function of [Fe/H]. The Sextans stars measured in this study are shown by diamonds with error bars; open diamonds are our target stars observed by Subaru HDS while filled diamonds are re-analyzed stars taken from the VLT archive. Previously studied Sextans stars are depicted by circles \citep{shetrone01, waoki09}. The abundances of Galactic halo stars are shown by asterisks taken from the SAGA database \citep{suda08}. Blue symbols show RGB stars that have similar metallicity ($-3.22 <$ [Fe/H] $< -2.65$) to the five stars. The right panel shows the histogram of these halo RGB stars. The histogram range shaded in blue shows the range where our five Sextans dwarf galaxy stars lie.}
         \label{fig:srfe}
   \end{figure*}
It should be noted that the treatment of scattering in the
  opacity calculations in the spectrum synthesis code is discussed in
  \citet{tafel10} in detail. We presume that this is not the
  reason for the discrepancy of Sr (and Ba) abundances because (1)
  scattering is included in the opacity calculation in our analysis;
  (2) no clear dependence of derived Fe abundances on the wavelengths
  of spectral lines is found in the analysis for HD~88609 (as shown in Fig.~\ref{fig:logewavel}), in contrast
  to the dependence found by \citet{tafel10}, where these latter authors do not include
  the effect of scattering; and (3) the Sr abundance derived by our
  analysis is {lower} than those of \citet{tafel10}, whereas 
    higher abundances are expected if the effect of scattering is not
  included according to their inspection.

The Sr abundance derived for the comparison star HD~88609 in our
  analysis is in good agreement with those of \citet{honda04} and \citet{honda07}. The differences
  between their results and ours are smaller
  than 0.1~dex. \citet{hansen12} obtained a higher Sr abundance for
  this object by about 0.3~dex. This is at least partially
  explained by the difference of microturbulent velocity: our value is
  0.7~kms$^{-1}$ larger than that of these latter authors, which could result in an abundance that is lower by  about
0.3~dex.

Although there remains a discrepancy between our results for S24-72 and S11-04 and those of Tafelmeyer et al. (2010), we expect that our results can be combined with those of the other three stars to which the same analysis technique is applied.
A comparison of [Sr/Fe] in the red giant stars of the Galactic halo is shown in Figure~\ref{fig:srfe}. Sr is clearly under-abundant in these stars in the Sextans dwarf galaxy compared to the average abundance ratio of field stars. The possible offset between our results and those of previous studies should be noted, taking into account the discrepancy in  derived Sr abundance between our analysis and that of \citet{tafel10}. Nevertheless, the possible offset is at the level of 0.4~dex, which does not affect the above result showing that Sr is under-abundant in the Sextans stars.

We obtained the abundance of Ba whilst taking the effect of hyperfine splitting into consideration \citep{mcwill98}. \citet{waoki09} measured the abundance of Ba from two lines at 4934 Å and 6141 Å. These latter authors did not use the line at 4554 Å because it is affected by a bad column on the CCD. 
 \citet{shetrone01} measured the  Ba abundance of S~49 from two lines at 5853 and 6141 Å without considering the effect of hyperfine splitting. 
 We measured the  Ba abundance of S~24-72 and S~11-04 using 4934, 6141, and 6496 Å, including the effect of hyperfine splitting. The differences in [Ba/Fe] and [Sr/Ba] abundance ratios between ours and  previous studies are given in Table~\ref{tab:srba}. The [Ba/Fe] of S~24-72 and S~11-04  is compared with other stars in the Sextans dwarf galaxy and halo stars in Figure~\ref{fig:srfe}. Ba is also under-abundant in these stars in the Sextans dwarf galaxy. 

\citet{tafel10} showed that  S~24-72 and S~11-04  have similar [Sr/Ba] abundance ratios ($\sim$0.8~dex). Since we obtained a lower Sr abundance, the Sr/Ba abundance ratio we measure is lower than that of \citet{tafel10}. We adopt the result obtained by our analysis which was also applied to the abundance measurements for S~10-14, S~11-13, and S~49 observed in the present work. The standard deviations for [Sr/Fe] and [Ba/Fe] for the five Sextans stars analyzed here are 0.27 and 0.25~dex, respectively. These values are similar to the errors estimated for [Sr/Fe] and [Ba/Fe], which are 0.26 and 0.27~dex,
on average, respectively. This indicates that we find no intrinsic scatter in these abundance ratios for the five stars studied here. The upper limit of the scatter would be comparable to the average errors estimated. As for [Sr/Ba], the standard deviation of the five Sextans stars is 0.07~dex, while the estimated error is 0.32~dex. Clearly, no intrinsic scatter of the abundance ratios is found in our results. The fact that the scatter of [Sr/Ba] is smaller than the errors possibly indicates that the errors are overestimated in our study, but we can still conclude that intrinsic scatter is not found.
   \begin{figure*}
   \centering
\includegraphics[width=8.0cm, angle=270]{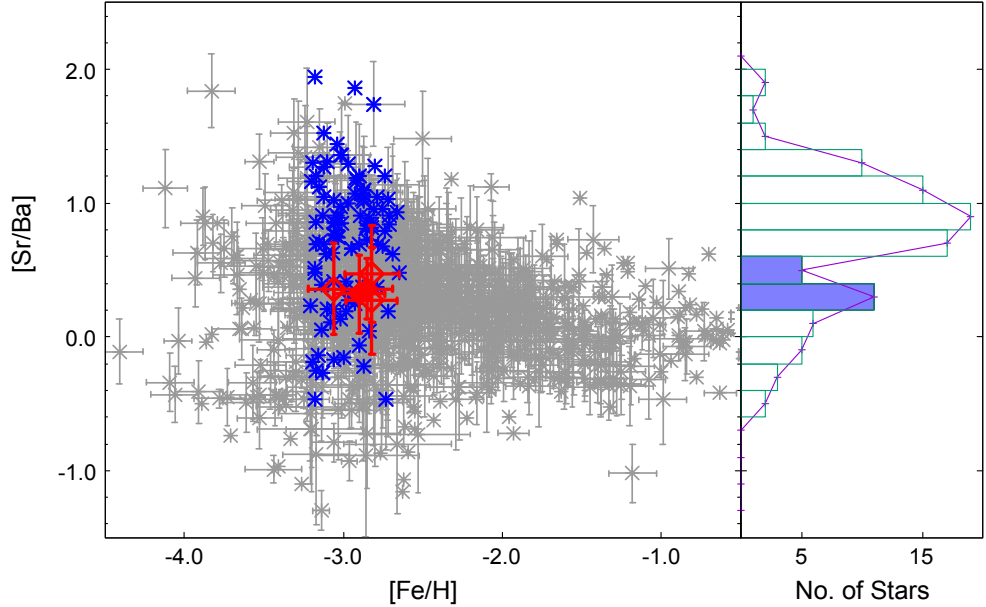}
      \caption{[Sr/Ba] as a function of [Fe/H] and [Ba/H]. The Sextans stars measured in this study are shown by diamonds with error bars; open diamonds are our target stars observed by Subaru HDS while filled diamonds are re-analyzed stars taken from the VLT archive. Blue symbols show 98 halo RGB stars that have similar metallicity ($-3.22 <$ [Fe/H] $< -2.65$) and [Ba/Fe] ($-1.95 <$[Ba/Fe] $< -0.77$) to the five stars. The right panel shows the histogram of [Sr/Ba] for the 98 halo stars. The range of the ratio (0.2< [Sr/Ba] <0.6) is shaded in blue, in which we see clumping of [Sr/Ba] with our five Sextans dwarf galaxy stars.}
         \label{fig:srba}
   \end{figure*}
\begin{table*}
\caption{Abundance ratios of our work and comparison with previous studies}
\label{tab:srba}
\centering

\begin{tabular}{lcccccccc}
\hline\hline
    \noalign{\smallskip}
Star&[Sr/Fe]&[Ba/Fe]&[Sr/Ba]&[Ba/H]&$\Delta$[Sr/Fe]&$\Delta$[Ba/Fe]&$\Delta$ [Sr/Ba]&Ref.\\
\hline
            \noalign{\smallskip}
    S~10-14 &$-1.05\pm0.28$&$-1.52\pm0.30$&$0.47\pm0.36$&$-4.34\pm0.23$&. . .& $-0.18$&. . . &(1)\\
    S~11-13 &$-1.39\pm0.33$&$-1.66\pm0.29$& $0.27\pm0.40$&$-4.48\pm0.21$&. . . &$-0.32$&. . .&(1) \\
    S~49&$-0.89\pm0.24$&$-1.25\pm0.32$& $0.36\pm0.34$&$-4.31\pm0.25$&. . .&$-0.20$&. . .& (2)\\
\hline
            \noalign{\smallskip}
S~24-72 &$-1.00\pm0.26$&$-1.32\pm0.23$&$0.32\pm0.29$&$-4.22\pm0.13$&$-0.79$&$-0.22$& $-0.57$&(3)\\
S~11-04 &$-0.64\pm0.19$&$-1.00\pm0.23$&$0.36\pm0.23$&$-3.85\pm0.12$&$-0.63$&$-0.15$&$-0.48$&(3)\\ 
\hline
            \noalign{\smallskip}
HD~88609 &$-0.20\pm0.20$&$-1.03\pm0.22$&$0.83\pm0.22$&$-4.00\pm0.09$&$-0.15$&$-0.22$& $0.07$&(4)\\

\hline
\end{tabular}
\tablefoot{The difference is taken as our results minus those of other works.\\
\tablebib{(1)~\citet{waoki09}; (2)~\citet{shetrone01}; (3)~\citet{tafel10}; (4)~\citet{honda07}
}}
\end{table*}

As we point out in Sect.1, abundance ratios of neutron-capture elements (e.g., [Sr/Ba]) confer an advantage for chemical tagging, because their abundance ratio could be sensitive to the enrichment by preceding nucleosynthesis events, which results in a large scatter in metal-poor stars in the Milky Way halo. Figure~\ref{fig:srba} shows the relation between [Sr/Ba] and metallicity of our five Sextans dwarf galaxy stars and the stars in the Milky Way halo which show a large scatter. On the contrary, the five Sextans dwarf galaxy stars show very good agreement in the [Sr/Ba] ratio. In the following section, we discuss the possible reasons for the clustering of [Sr/Ba].
   \begin{figure}
   \centering
\includegraphics[width=9cm]{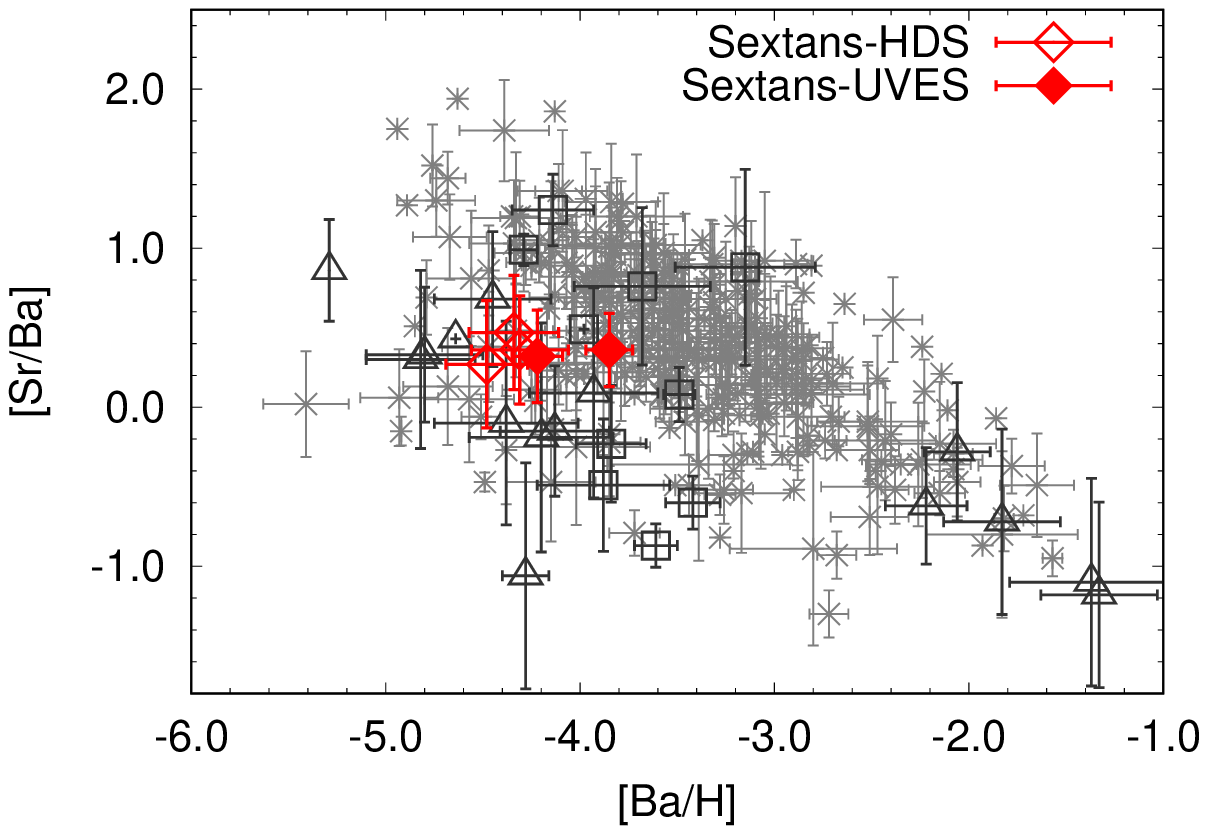}
      \caption{[Sr/Ba] as a function of [Ba/H], showing a comparison between our results and those of previous studies. Stars in classical dwarf spheroidal galaxies are shown by squares for Carina \citep{venn12}, Draco \citep{cohen09}, Sculptor \citep{starkenburg13, jablonka15}, and Ursa Minor \citep{kirby12, cohen10}, and stars in ultra faint dwarf galaxies are shown by triangles for Reticulum II \citep{ji16}, Coma Berenices \citep{frebel10}, Leo IV \citep{simon10}, Triangulum II \citep{kirby17}, Tucana II \citep{chiti18}, Horologium I \citep{nagasawa18}, Bootes I \citep{ishigaki14} and Ursa Major II \citep{frebel10}. The asterisks show Milky Way halo stars. The stars in dwarf galaxies and Milky Way halo stars shown here have [Fe/H] between $-3.22$ and $-2.65$. }
         \label{fig:BaHdwarfs}
   \end{figure}

\section{Discussion}
\subsection{Clustering of abundance ratios}
The abundance analysis for five extremely metal-poor stars in the Sextans dwarf galaxy confirms the clustering of abundances of Mg, Ca, and Ba as found by previous studies \citep{waoki09, tafel10}.  Here we focus on the abundance ratios of Sr and Ba, which could provide a new examination of the clustering of chemical abundance ratios in this galaxy.

The blue stars in Figure \ref{fig:srfe} are halo stars with $-3.22<$[Fe/H]$<-2.65$, which covers the metallicities of the five Sextans stars. The right panels of Figure \ref{fig:srfe} show the distributions of [Sr/Fe] and [Ba/Fe] for these halo stars. The regions corresponding to the [Sr/Fe] and [Ba/Fe] values of the five Sextans stars are shaded in blue. 

Figure~\ref{fig:srba} shows that the five stars studied by the present work have similar [Sr/Ba] ratios. This clumping is distinctive compared to the [Sr/Ba] spread seen in the halo stars. The blue symbols in Figure~\ref{fig:srba} show 98 halo RGB stars that have similar metallicity and [Ba/Fe] ratio to the five target stars ($-3.22 <$ [Fe/H] $< -2.65$ and $-1.95 <$[Ba/Fe] $< -0.77$), showing a very wide spread of [Sr/Ba] from $-0.6$ up to 2.0. The stars are not uniformly spread, as can be seen in the histogram showing the distribution of [Sr/Ba] of those 98 halo RGB stars, with the peak around [Sr/Ba]$\sim$0.9. 
Interestingly, our five stars do not clump at the peak but at lower [Sr/Ba] regions. The five stars lie in the range 0.2$<$[Sr/Ba]$<$0.6 (highlighted bars in the histogram, Figure~\ref{fig:srba}). Within this range, there are 16 RGB halo stars out of the selected 98 stars. If the five stars of our sample are assumed to have a similar distribution of [Sr/Ba] to that of the field halo stars with similar [Ba/Fe], the probability that the five Sextans metal-poor stars clump in the range of 0.2$<$[Sr/Ba]$<$0.6 is $ \left(16/98\right)^{5-1}=0.07\%$. If we adopt a wider range, namely $-0.2<$[Sr/Ba]$<$0.8, the number of RGB stars in the range is 44, and the probability estimated in the same manner is 4.06$\%$. This result indicates that the clustering of the [Sr/Ba] abundance ratios of the five stars is significant, which is not explained if the [Sr/Ba] distribution of halo stars is assumed. Similar estimates of the probabilities that the five Sextans stars clump in the ranges given above for [Sr/Fe] and [Ba/Fe] are as low as 0.1$\%$. It should be noted that the [Sr/Fe] and [Ba/Fe] values of very metal-poor stars in dwarf galaxies are relatively low in general, and therefore it might not be meaningful to calculate the probability assuming the distribution of these abundance ratios of halo stars. By contrast, the [Sr/Ba] values of very metal-poor stars in dwarf galaxies show a wide distribution (see below).

The clustering of [Sr/Ba] could be evidence that the five metal-poor stars in the Sextans dwarf galaxy were formed in an environment of homogeneous chemical composition. Our result does not prove that all the extremely metal-poor stars in this galaxy were formed in an environment of homogenous composition since there is at least one other star studied by \citet{waoki09} with similar metallicity but with higher Mg and Ba abundance (Figure \ref{mgfe} and \ref{fig:srba}). There could be more than one cluster in this dwarf galaxy, but this has not yet been confirmed. However, the fact that a group of extremely metal-poor stars in the Sextans dwarf galaxy share a similar [Sr/Ba] ratio at least provides constraints on the numerical simulation of possible hierarchical formation and merger history of the Sextans dwarf galaxy.
 
 Such clustering of chemical abundance ratios in very metal-poor stars is predicted by models of dwarf galaxy formation \cite[e.g.,][]{bland10}, which is in clear contrast to the dispersion in field halo stars. The large and smooth distribution in abundance ratios for elements in the field halo stars could be the result of combining a large number of clusters. More samples of clustering in elemental abundances for metal-poor stars in dwarf galaxies would provide a means to strongly constrain the scenario for Milky Way formation.
 
\begin{table}
\caption{[Sr/Ba] and [Ba/H] of stars in dwarf galaxies of Fig. \ref{fig:srba}}
\label{tab:BaH-dwarfs}
\centering
\begin{tabular}{lrcc}
\hline\hline
            \noalign{\smallskip}
 \multicolumn{4}{c}{Classical dwarf spheroidal galaxies}  \\
Star&[Sr/Ba]&[Ba/H]&Ref.\\
\hline
            \noalign{\smallskip}
Car-1087&$-0.49$&$-3.88$&(1)\\
Car-7002&$-0.23$&$-3.84$&(1)\\
\hline
            \noalign{\smallskip}
19219&$-0.87$&$-3.61$&(2)\\
19629&$-0.60$&$-3.42$&(2)\\
\hline
            \noalign{\smallskip}
Scl\_03\_170&$0.76$&$-3.68$&(3)\\
Scl024\_01&0.88&$-3.15$&(3)\\
scl\_03\_059&0.08&$-3.49$&(4)\\
\hline
            \noalign{\smallskip}
UMiJI19&$0.99$&$-4.29$&(5)\\
UMi33533&1.24&$-4.14$&(5)\\
UMi20103&0.49&$-3.98$&(6)\\
\hline \hline
            \noalign{\smallskip}
 \multicolumn{3}{c}{Ultra faint dwarf galaxies} &  \\
Star&[Sr/Ba]&[Ba/H]&Ref.\\
\hline
            \noalign{\smallskip}
DES\_J033454-540558&$-1.10$&$-1.37$&(7)\\
DES\_J033447-540525&$-0.72$&$-1.83$&(7)\\
DES\_J033523-540407&$-0.62$&$-2.22$&(7)\\
DES\_J033537-540401&$-1.18$&$-1.33$&(7)\\
DES\_J033607-540235&$-0.28$&$-2.06$&(7)\\
\hline
            \noalign{\smallskip}
ComBer-S2&$0.33$&$-4.80$&(8)\\
\hline
            \noalign{\smallskip}
LeoIV-S1&$0.43$&$-4.64$&(9)\\
\hline
            \noalign{\smallskip}
40&$0.86$&$-5.29$&(10)\\
\hline
            \noalign{\smallskip}
TucII-06&$-0.19$&$-4.20$&(11)\\
TucII-011&$0.30$&$-4.82$&(11)\\
TucII-203&$-0.10$&$-4.38$&(11)\\
TucII-033&$-0.15$&$-4.13$&(11)\\
\hline
            \noalign{\smallskip}
DESJ025535-540643&$0.09$&$-3.93$&(12)\\
\hline
            \noalign{\smallskip}
Boo-094&$-1.06$&$-4.28$&(13)\\
\hline
            \noalign{\smallskip}
UMaII-S1&$0.68$&$-4.45$&(8)\\
\hline
   \hline
\end{tabular}
\tablebib{(1)~\citet{venn12}; (2)~\citet{cohen09}; (3)~\citet{starkenburg13}; (4)~\citet{jablonka15}; (5)~\citet{cohen10}; (6)~\citet{kirby12}; (7)~\citet{ji16}; (8)~\citet{frebel10}; (9)~\citet{simon10}; (10)~\citet{kirby17}; (11)~\citet{chiti18}; (12)~\citet{nagasawa18}; (13)~\citet{ishigaki14}
}
\end{table}

\subsection{Sr and Ba abundances in dwarf galaxies}
Figure~\ref{fig:BaHdwarfs} shows the relation between [Sr/Ba] and [Ba/H] compared with Milky Way halo stars with metallicity $-3.22<$[Fe/H]$<-2.65$ and stars in other previously studied dwarf spheroidal galaxies in the same metallicity range (Table \ref{tab:BaH-dwarfs}). As mentioned by \citet{tafel10} , \citet{francois07} described a strong anti-correlation between [Sr/Ba] and [Ba/H]. In the Milky Way halo, [Sr/Ba] increases from a solar abundance ratio to $1.0\pm0.4$~dex as [Ba/H] decreases from $-3$ to $-4$~dex. Below [Ba/H]=$-5$, there are fewer samples of stars to show the trend, and other stars show lower abundance ratios. Some objects in dwarf galaxies follow the anti-correlation trend, but many others show lower [Sr/Ba] than the trend. We note that within the same classical dwarf spheroidal galaxies, there is generally good agreement between stars in terms of  [Ba/H], but the number of stars studied for each galaxy is still too small. Our measurements of Sr and Ba abundances for five stars in Sextans are a first example of clustering of the abundance ratios including neutron-capture elements. 

The Milky Way halo stars in Figure~\ref{fig:BaHdwarfs} show not only the strong anti-correlation but also a branch with [Sr/Ba]$\sim 0$, which starts to extend at [Ba/H]$\sim$$-4$. This is also apparent in the dwarf galaxy stars, including the Sextans dwarf galaxy stars we analyzed. This branch is also mentioned by \citet{mashon17}. These latter authors suggest that this branch is due to a second producer of Sr which is independent of the production of Ba and operated below [Ba/H] $\sim$$-$4. It is also possible that the Sr and Ba of the stars that lie on this branch come from the same origin. Interestingly, we can see that stars in ultra-faint dwarf galaxies other than Reticulum II lie near this branch. Reticulum II, an ultra-faint dwarf galaxy known to have enriched r-process element abundance also shows clustering in Figure~\ref{fig:BaHdwarfs}. In this figure, we selected those stars that have similar metallicity to our sample stars. However, other stars in Reticulum II that show a wide metallicity distribution show a small dispersion of [Sr/Ba] \citep{ji16}. Given the wide metallicity distribution, the homogeneity of neutron-capture element ratios found in Reticulum II could be due to different mechanisms from those that lead to the clustering found in dwarf spheroidal galaxies.This homogeneity could be due to a single r-process event that has polluted the whole progenitor of the ultra-faint dwarf galaxy.

The classical dwarf spheroidal stars, in general, have slightly higher [Ba/H] than the majority of those in ultra-faint dwarf galaxies. The stars with [Sr/Ba]$>$ 0.0 (except for the Sextans dwarf galaxy stars) lie on top of the thick anti-correlation trend of the Milky Way halo stars. On the other hand, classical dwarf spheroidal stars with negative [Sr/Ba] (two stars in Carina and two in Draco) do not follow the anti-correlation trend and are not found in the aforementioned branch. We note that within the same dwarf galaxies, some stars have a similar Sr/Ba ratio, but the number of stars studied is too small for a meaningful analysis of clustering. On the other hand, Tucana II, an ultra faint dwarf galaxy that has four stars in our range of [Fe/H] shows moderate clustering in [Sr/Ba]. The majority of these stars are located in the branch mentioned in the previous paragraph.

Overall, the stars in classical dwarf spheroidal galaxies do not show the clear anti-correlation trend in [Sr/Ba] as a function of [Ba/H], which we see clearly in the Milky Way halo stars. On the other hand, the abundance ratios of the ultra-faint dwarf galaxy stars generally show an anti-correlation trend with smaller inclination, from the lower part of the anti-correlation on the Milky Way halo stars at higher [Ba/H] to the ``branch'' at [Sr/Ba]$\sim 0$ at lower [Ba/H].


\section{Summary}
We analyzed the high-resolution spectra of five metal-poor red giants ([Fe/H]$< -$2.8) in the Sextans dwarf spheroidal galaxy. Three of the stars were observed with Subaru HDS, while two stars were taken from the VLT archive. The abundances of eight chemical elements were measured. The abundance of Sr was measured for the first time in the three targets observed with Subaru. 

We have confirmed that, in general, the $\alpha$/Fe abundance ratios of dwarf galaxies are slightly lower than the average for stars in the Galactic halo. The abundance ratios of Mg, Ca, and Ba for the five stars of our sample show good agreement with one another. 

The Sr/Ba abundance ratios of the five metal-poor stars are also in good agreement with each other. The clumping is distinctive compared to the [Sr/Ba] spread seen in the halo stars with similar metallicity. The probability of such clumping of [Sr/Ba] is very small if the [Sr/Ba] distribution of halo stars is assumed. The clustering of [Sr/Ba] that we observe is good evidence that these stars were formed in an environment of homogeneous chemical composition.

Previous studies of other dwarf spheroidal galaxies, including ultra-faint dwarf galaxies generally, reveal two general trends for the Milky Way halo stars. One is the [Ba/H] versus [Sr/Ba] anti-correlation trend, and the other is the branch of stars at [Ba/H]$\sim$$-4$, with a flatter trend of [Sr/Ba]. Our results provide constraints on the possible different nucleosynthesis origins for Sr, which starts at [Ba/H]$\sim$$-4$. Sextans dwarf galaxy stars lie at the start of the branch, which could provide a hint to the formation and chemical origin of the Sextans dwarf galaxy.

Our results also provide a constraint on the formation and chemical evolution of the Sextans dwarf galaxy and provide clues as to the roles of dwarf galaxies as building blocks for making large structures. Further surveys of metal-poor stars in dwarf spheroidal galaxies, along with their kinematic measurements, will improve our understanding of the role of dwarf galaxies in the evolution of the Milky way. After this work was completed, to further chemical abundance studies of the Sextans dwarf galaxy were submitted \citep[submitted]{theler19, lucchesi20}. This dwarf galaxy may be a particularly interesting future target as a relatively large number of its very metal-poor stars have been observed.


\begin{acknowledgements}\\
This research has made use of the SIMBAD database, operated at CDS, Strasbourg, France.
MA acknowledges support from the IMPRS on Astrophysics at the LMU M{\"u}nchen. PFR acknowledges support from the ``Action F{\'e}d{\'e}ratrice Etoiles'' of the Paris-Meudon Observatory. We express our sincere gratitude to Dr. Yuhri Ishimaru, who, although no longer with us, continues to inspire us with her works. The early phase of this study is based on discussions with her.
\end{acknowledgements}


%
%
\bibliographystyle{aa}
\bibliography{Sextans-paper-ref-Finalcorr-arxiv}

\begin{appendix}
\section{Line Data and Equivalent Widths}
   \longtab[1]{
          \begin{longtable}{lccrrrp{0.05pt}rrp{0.05pt}rr}
           \caption{Line Data and Equivalent Widths}\\
                             \label{tab:linelist}\\
            \hline
            \noalign{\smallskip}
& & & &\multicolumn{2}{c}{S$10-14$} & & \multicolumn{2}{c}{S$11-13$} & & \multicolumn{2}{c}{S 49}\\
\cline{5-6}\cline{8-9}\cline{11-12}\\
Elem.& $\lambda$& L.E.P& log {\it gf}& log $\epsilon$&EW& &log $\epsilon$ & EW& & log $\epsilon$&EW\\
&(\AA) &(eV) &  &  & (m\AA)& & &(m\AA) & & & (m\AA)\\
\hline
\endfirsthead
\caption{continued.}\\
 \hline
            \noalign{\smallskip}
& & & &\multicolumn{2}{c}{S$10-14$} & & \multicolumn{2}{c}{S$11-13$} & & \multicolumn{2}{c}{S 49}\\
\cline{5-6}\cline{8-9}\cline{11-12}\\
Elem.& $\lambda$& L.E.P& log {\it gf}& log $\epsilon$&EW& &log $\epsilon$ & EW& & log $\epsilon$&EW\\
&(\AA) &(eV) &  &  & (m\AA)& & &(m\AA) & & & (m\AA)\\
            \hline
            \noalign{\smallskip}
\endhead
\hline
\endfoot
            \noalign{\smallskip}


Mg {\tiny I}&4571.10& 0.00 & $-5.69$
& 4.98& 55.1 &
& 5.03&77.2&
& 4.79 & 67.0\\		

Mg {\tiny I}& 4702.99& 4.33 & $-0.44$
& . . . & . . .&
& . . . & . . .&
& 4.97 & 48.5\\

Mg {\tiny I}&5183.60& 2.72 & $-0.24$
& 4.64& 179.4&
& 4.71&187.9&
& 4.51 & 186.5\\	

Mg {\tiny I}& 5528.40 &4.35 & $-0.50$
& 4.92 & 42.5&
& 5.05 & 56.2&
& . . . & . . . \\

Ca {\tiny I}&4454.78&1.90 & $0.26$
& . . . & . . . &
& . . . & . . . &
& 3.63& 87.4 \\

Ca {\tiny I}&4455.89&1.90 & $-0.53$
& . . . & . . . &
& 3.61& 35.9&
& . . . & . . . \\
 
Ca {\tiny I}&5265.56&2.52 & $-0.11$ 
& . . . & . . . &
& 3.69 & 25.4&
& 3.57 & 21.5 \\

Ca {\tiny I}&5588.76&2.53 & 0.36 
& 3.65 & 34.8 &
& 3.79& 49.1 &
& 3.52& 37.1 \\


Sc {\tiny II}& 4314.10& 0.62 & 0.10
& . . . & . . . &
& 0.12 & 112.0&
& 0.05& 102.7\\

Sc {\tiny II}& 4374.46& 0.62 & $-0.42$
& . . . & . . .&
& 0.17 & 85.8&
& $-0.09$ & 76.8 \\

Sc {\tiny II}& 4400.40& 0.61 & $-0.54$
& 0.05 & 142.1 &
& 0.00 & 81.7&
& . . . & . . .\\

Sc {\tiny II}& 4415.56& 0.60 & $-0.67$
& . . . & . . .&
& 0.17 & 75.6&
& . . . & . . . \\

Ti {\tiny I}& 4533.24& 0.84 &0.54
 & $<$2.21 & $<$52.3 &
 & . . . & . . . &
& 1.72 &  39.2  \\

Ti {\tiny I}& 4656.47& 0.00 &$-1.29$
 & $<$2.58 &$<$27.8&
 & . . . & . . . &
& . . . & . . . \\

Ti {\tiny I}& 4981.73& 0.85 &0.57
 & . . . & . . . &
& 2.26& 72.2&
& . . . & . . . \\

Ti {\tiny I}& 4991.07& 0.84 &0.45
 & $<$2.02& $<$38.3 &
& 2.10 & 55.2&
& 1.82& 42.6\\

Ti {\tiny I}& 5036.46& 0.19 &$1.44$
& . . . & . . . &
 & . . . & . . . &
& $<$2.33& $<$17.6\\

Ti {\tiny I}& 5210.38& 0.05 &$-0.82$
 & . . . & . . . &
& 2.13& 42.3&
& . . . & . . . \\

Ti {\tiny II}&4025.13& 0.61 & $-2.11$ 
& . . . & . . .  &
& 2.07 & 72.2&
& . . . & . . .\\	

Ti {\tiny II}&4290.22& 1.17 & $-0.87$ 
& . . . & . . .  &
& 2.10 & 102.0&
& . . . & . . . \\	



Ti {\tiny II}&4395.84 & 1.24 & $-1.93$ 
& . . . & . . . &
& 2.05 & 38.1 &
& 2.06 & 41.5  \\

Ti {\tiny II}&4417.72 & 1.17 & $-1.19$
& 2.16& 91.5 &
& 2.45 &105.7&
& 1.74& 75.0\\

Ti {\tiny II}&4443.80 & 1.08 & $-0.71$
& . . . & . . .  &
& 1.97 & 111.4 &
& 1.77 & 110.4 \\

Ti {\tiny II}&4464.45& 1.16 & $-1.81$ 
& . . . & . . .  &
& 2.20 & 60.6&
& . . . & . . .\\	

Ti {\tiny II}&4468.49&1.13& $-0.63$ 
& . . . & . . .  &
& 2.05 & 116.2&
& . . . & . . .\\	

Ti {\tiny II}&4533.97& 1.24 & $-0.77$ 
& . . . & . . .  &
& 1.99 & 101.3&
& . . . & . . .\\	

Ti {\tiny II}&4563.77& 1.22 & $-0.96$ 
& 2.13& 100.8&
& 2.33& 110.0&
& 2.14& 109.2 \\	

Ti {\tiny II}&4571.97& 1.57 &$ -0.31$ 
& 1.80& 94.0&
& 2.54& 129.7 &
& 2.18& 121.1\\

Ti {\tiny II}& 5129.16& 1.88 &$-1.34$
& . . . & . . .  &
& 2.20 & 37.7&
& . . . & . . .\\	

Ti {\tiny II}&5185.90& 1.89& $-1.41$ 
& 2.32 & 35.6 &
& 2.12& 29.2&
& 2.27& 39.5 \\

Ti {\tiny II}& 5226.54& 1.57 & $-1.23$ 
& 2.09 & 58.5 &
&2.16& 68.7& 
& 2.09& 68.9\\

Ti {\tiny II}& 5336.79& 1.58 & $-1.60$ 
& . . . & . . .  &
& 2.20 &46.2 &
& . . . & . . .\\	

Cr {\tiny I}& 4254.33& 0.00 & $-0.09$ 
&2.09 & 111.5 &
&2.18 & 120.8&
& 1.97& 118.8 \\

Cr {\tiny I}& 4289.72& 0.00 & $-0.36$ 
&. . . & . . .& 
&2.20  & 112.2&
&. . . & . . . \\

Cr {\tiny I}& 5345.80& 1.00 & $-0.95$ 
&2.60& 27.4& 
& 2.38& 28.1&
& 2.15 & 29.2\\	

Cr {\tiny I}& 5409.77& 1.03 & $-0.67$ 
&2.60 & 41.8& 
& 2.43& 44.3&
& 2.30 & 38.5 \\

Mn {\tiny I}& 4048.99& 2.16 & $-0.13$ 
&. . .&. . .&
&2.15&100.0&
&. . . &. . . \\

Mn {\tiny I}& 5407.42& 2.14 & $-1.74$ 
&2.73 & 27.7& 
&. . .&. . .&
&. . . &. . . \\	

Mn {\tiny I}& 5420.36& 2.14 & $-1.46$ 
&2.69 &41.3& 
&. . .&. . .&
&. . . &. . . \\			

Fe {\tiny I}& 4132.90 & 2.85 & $-1.01$ 
& 4.66 & 41.3 &
&. . . & . . .& 
& . . . & . . . \\

Fe {\tiny I}& 4143.42 & 3.05 & $-0.20$ 
&. . . & . . .& 
&4.22 & 55.2& 
& . . . & . . . \\

Fe {\tiny I}& 4143.87 & 1.56 & $-0.51$ 
&. . . & . . .& 
&3.81 & 109.7& 
& . . . & . . . \\

Fe {\tiny I}& 4152.17& 0.96 & $-3.23$ 
& 5.11 & 72.5 &
&. . . & . . .& 
& . . . & . . . \\

Fe {\tiny I}& 4187.80& 2.43 & $-0.55$ 
& 4.69 & 97.3 &
&. . . & . . .& 
& . . . & . . . \\

Fe {\tiny I}& 4206.70 & 0.05 & $-3.96$ 
&. . . & . . .& 
& 4.87& 92.7& 
& . . . & . . . \\

Fe {\tiny I}& 4216.18 & 0.00 & $-3.36$ 
&. . . & . . .& 
& 4.54& 108.9& 
& . . . & . . . \\

Fe {\tiny I}& 4235.94& 2.43 & $-0.34$ 
& 4.29& 88.5 &
&4.49&99.8& 
& . . . & . . . \\

Fe {\tiny I}& 4250.12 & 2.47 & $-0.41$ 
&. . . & . . .& 
&4.15 & 79.2& 
& . . . & . . . \\

Fe {\tiny I}& 4250.79 &1.56 & $-0.71$ 
&. . . & . . .& 
&4.46 & 132.5& 
& . . . & . . . \\

Fe {\tiny I}& 4271.15& 2.45 & $-0.34$ 
& 4.44 & 95.6 &
&. . . & . . .& 
& 4.09 & 88.2\\

Fe {\tiny I}& 4282.40 & 2.18 & $-0.78$ 
&. . . & . . .& 
&4.83& 110.6& 
& . . . & . . . \\

Fe {\tiny I}& 4299.24 & 2.43 & $-0.38$ 
&. . . & . . .& 
&4.48 & 98.7& 
& . . . & . . . \\

Fe {\tiny I}& 4337.05& 1.56 & $-1.70$ 
&. . . & . . .& 
&4.73 & 101.4& 
& . . . & . . . \\

Fe {\tiny I}& 4375.93& 0.00 & $-3.02$ 
& 4.03 & 93.1&
&4.56 & . 129.6& 
& 4.10 & 115.9\\

Fe {\tiny I}& 4427.31& 0.05 & $-2.92$ 
& 4.58 & 122.8 &
&4.68&138.1& 
& . . . & . . . \\

Fe {\tiny I}& 4442.34 & 2.20 & $-1.26$ 
&. . . & . . .& 
&4.56 & 76.6& 
& . . . & . . . \\

Fe {\tiny I}& 4443.19 & 2.86 & $-1.04$ 
&. . . & . . .& 
&4.48 & 38.1& 
& 4.33 & 33.5\\

Fe {\tiny I}& 4447.72 & 2.22 & $-1.34$ 
&. . . & . . .& 
&4.33 & 57.9& 
& 4.58& 78.2\\

Fe {\tiny I}& 4461.65& 0.09 & $-3.21$ 
& 4.27 & 91.3&
&.4.65 & 120.3& 
& . . . & . . . \\

Fe {\tiny I}& 4489.74& 0.12 & $-3.97$ 
& 5.00& 87.7 &
&5.04 & 100.7& 
& . . . & . . . \\

Fe {\tiny I}& 4494.56 & 2.20 & $-1.14$ 
&. . . & . . .& 
&4.87 & 98.0& 
& 4.25 & 73.6\\

Fe {\tiny I}& 4528.61 & 2.18 & $-0.82$ 
&. . . & . . .& 
&4.23 &84.2& 
& 4.17 & 88.5\\

Fe {\tiny I}& 4531.15& 1.49 & $-2.16$ 
&. . . & . . .& 
&5.04 & 102.1& 
& . . . & . . . \\

Fe {\tiny I}& 4592.65& 1.56 & $-2.45$ 
&. . . & . . .& 
&. . . & . . .& 
& 4.65& 68.9\\

Fe {\tiny I}& 4602.94& 1.49 & $-2.21$ 
& 4.54 & 65.0 &
&. . . & . . .& 
& 4.30& 67.6\\

Fe {\tiny I}& 4859.74& 2.88 & $-0.76$ 
& 4.84 &69.9 &
& 4.42 & 51.6& 
& 4.17 & 40.5\\

Fe {\tiny I}& 4871.32& 2.87 & $-0.36$ 
& 4.46 & 72.4 &
&. . . & . . .& 
& . . . & . . . \\

Fe {\tiny I}& 4872.14 & 2.88 & $-0.57$ 
&. . . & . . .& 
&4.64& 74.9& 
& 4.47 & 71.4\\

Fe {\tiny I}& 4890.76& 2.88 & $-0.39$ 
& 4.45& 69.1 &
&. . . & . . .& 
& . . . & . . . \\

Fe {\tiny I}& 4891.49& 2.85 & $-0.11$ 
& 4.11 & 67.5 &
&. . . & . . .& 
& 4.03 &75.1\\

Fe {\tiny I}& 4918.99& 2.87 & $-0.34$ 
& 4.38 & 68.8 &
& 4.34 & 72.1& 
& 4.30 & 76.7\\

Fe {\tiny I}& 4924.77 & 2.28 & $-2.26$ 
&. . . & . . .& 
&4.88& 35.4& 
& . . . & . . . \\

Fe {\tiny I}& 4938.81 & 2.88 & $-1.08$ 
&. . . & . . .& 
&4.67 & 47.6& 
& . . . & . . . \\

Fe {\tiny I}& 4939.69& 0.86 & $-3.34$ 
& 5.13 & 82.5 &
&4.84 & 77.8& 
& 4.88 &87.6 \\

Fe {\tiny I}& 4966.09 & 3.33 & $-0.87$ 
&. . . & . . .& 
&4.88 & 39.3& 
& . . . & . . . \\

Fe {\tiny I}& 4994.13& 0.92 & $-2.96$ 
& 5.03 & 96.5 &
&4.91& 99.3& 
& . . . & . . . \\

Fe {\tiny I}& 5012.07& 0.86 & $-2.64$ 
& 4.65 & 96.8 &
&4.88& 118.3& 
& 4.55 & 110.2\\

Fe {\tiny I}& 5041.07& 0.96 & $-3.09$ 
& 4.85 & 76.3 &
&4.99& 94.1& 
& 4.67 & 83.3\\

Fe {\tiny I}& 5041.76 & 1.49 & $-2.20$ 
&. . . & . . .& 
&4.69 & 88.8& 
& 4.56 & 89.5\\

Fe {\tiny I}& 5049.82 & 2.28 & $-1.34$ 
&. . . & . . .& 
&4.54& 70.7& 
& 4.38 & 66.9\\

Fe {\tiny I}& 5051.64& 0.92 & $-2.80$ 
& 4.65 & 90.9 &
& 4.85& 105.5& 
& 4.70 & 106.8\\

Fe {\tiny I}& 5079.74 &0.99 & $-3.22$ 
&. . . & . . .& 
&4.96 &83.0& 
& . . . & . . . \\

Fe {\tiny I}& 5083.34& 0.96 & $-2.96$ 
& 4.96 & 89.8 &
&4.91& 97.3& 
& 4.63 & 89.6\\

Fe {\tiny I}&5098.70 & 2.18 & $-2.03$ 
&. . . & . . .& 
&4.84 & 55.6& 
& . . . & . . . \\

Fe {\tiny I}& 5123.72 & 1.01 & $-3.07$ 
&. . . & . . .& 
&4.92& 88.2& 
& . . . & . . . \\

Fe {\tiny I}& 5127.36& 0.92 & $-3.31$ 
& 5.08& 79.2 &
&5.26& 100.9& 
& 4.76 & 79.4\\

Fe {\tiny I}& 5142.93& 0.96 & $-3.08$ 
& 4.83 & 73.7 &
&4.84& 87.5& 
& . . . & . . . \\

Fe {\tiny I}& 5166.28& 0.00 & $-4.20$ 
&. . . & . . .& 
&. . . & . . .& 
& 4.75 & 99.8 \\

Fe {\tiny I}& 5194.94& 3.23 & $-2.22$ 
&. . . & . . .& 
&4.37 & 72.2& 
& 4.05 & 57.6\\

Fe {\tiny I}& 5216.27 & 1.61 & $-2.15$ 
&. . . & . . .& 
&4.81&90.9& 
& 4.53 & 82.3\\

Fe {\tiny I}& 5217.39& 3.21 & $-1.07$ 
&. . . & . . .& 
&. . . & . . .& 
& 4.64 & 27.8 \\

Fe {\tiny I}&5225.53 & 0.11 & $-4.79$ 
&. . . & . . .& 
&5.05 & 63.7& 
& . . . & . . . \\

Fe {\tiny I}& 5266.56& 3.00 & $-0.88$ 
& 4.68 & 77.2 &
&. . . & . . .& 
& . . . & . . . \\

Fe {\tiny I}& 5269.54 & 0.86 & $-1.32$ 
&. . . & . . .& 
&4.28 &163.2& 
& 4.12 &164.7\\

Fe {\tiny I}& 5281.79 & 3.04 & $-0.83$ 
&. . . & . . .& 
&4.48 & 42.1& 
& . . . & . . . \\

Fe {\tiny I}& 5283.63 & 3.24 & $-0.52$ 
&. . . & . . .& 
&4.64 & 53.8& 
& . . . & . . . \\

Fe {\tiny I}& 5324.18& 3.21 & $-0.10$ 
& 4.49& 66.4 &
&4.62&80.8& 
& . . . & . . . \\

Fe {\tiny I}& 5328.04 & 0.92 & $-1.47$ 
&. . . & . . .& 
& 4.39& 157.6& 
& . . . & . . . \\

Fe {\tiny I}& 5328.53 & 1.56 & $-1.85$ 
&. . . & . . .& 
&5.24& 135.1& 
& . . . & . . . \\

Fe {\tiny I}& 5339.93 & 3.27 & $-0.65$ 
&. . . & . . .& 
&4.72& 49.8& 
& . . . & . . . \\

Fe {\tiny I}& 5341.02& 1.61 & $-1.95$ 
&. . . & . . .& 
&4.74& 99.7& 
& . . . & . . . \\

Fe {\tiny I}& 5446.92 &0.99 & $-1.91$ 
&. . . & . . .& 
&4.36 & 126.8& 
& . . . & . . . \\

Fe {\tiny I}& 5455.61& 1.01 & $-2.10$ 
&. . . & . . .& 
&4.93 &147.2 &
& 4.54 & 135.5\\

Fe {\tiny I}& 5497.52& 1.01 & $-2.85$ 
& 4.89 & 92.1&
&. . . & . . .& 
& . . . & . . . \\

Fe {\tiny I}& 5501.47& 0.96 & $-3.05$ 
& 5.01& 90.7&
&5.16& 111.0& 
& 4.66 &90.0\\

Fe {\tiny I}& 5506.78& 0.99 & $-2.80$ 
&. 4.98& 102.07& 
&4.87 & 107.1& 
& 4.59 & 99.5\\

Fe {\tiny I}& 5615.64& 3.33 & $0.05$ 
& 4.45& 65.7&
&4.42& 71.0& 
& 4.33 & 70.7\\

Fe {\tiny II}& 4233.17 & 2.58 & $-1.97$ 
&. . . & . . .& 
&4.84&76.3&
& . . . & . . . \\

Fe {\tiny II}& 4416.83& 2.78 & $-2.54$ 
&. . . & . . .& 
&. . . & . . .& 
& 4.59& 40.3 \\

Fe {\tiny II}&4508.29& 2.86 & $-2.44$ 
&. . . & . . .& 
&5.09 &58.9 & 
& . . . & . . . \\

Fe {\tiny II}& 4923.93& 2.89 & $-1.26$ 
& 4.47 & 99.4 &
&4.70& 90.4& 
& 4.40 & 85.2\\

Fe {\tiny II}& 5018.45& 2.89 & $-1.10$ 
& 4.29 & 99.2 &
&. . . & . . .& 
& 4.68 & 104.4\\

Fe {\tiny II}& 5276.00& 3.20 & $-2.01$ 
& 4.39 & 33.8 &
&4.71& 46.9& 
& . . . & . . . \\
		
Ni {\tiny I}& 5476.90 & 1.83 &$-0.78$
&3.40& 76.4 &
&3.34 &82.9&
& 3.05& 70.5 \\

Zn {\tiny I}& 4810.53& 4.08 & $-0.17$
&. . .&. . .&
& 2.25 &35.7&
&2.02 &26.0\\

Sr {\tiny II}&4077.71& 0.00 & 0.17 
& $-1.08$ & 129.9 &
& . . .& . . .&
& $-1.04$ &161.9\\

Sr {\tiny II}&4215.52& 0.00 & $-0.17$
&$-0.92$ & 131.4 &
&$-1.34$ &117.8&
& $-0.80$ & 146.8 \\

Y {\tiny II}&4398.01& 0.13 & $-1.00$
&$<-0.96$&$<$25.3&
& $<-1.24$ &$<$20.3&
& . . .& . . .\\
	
Ba {\tiny II}& 4554.03& 0.00 & 0.14
&$-2.22$&96.8&
& $-2.35$ &97.0&
& . . .& . . .\\

Ba {\tiny II}& 4934.10& 0.00 & $-0.16$
&$-2.09$&88.0&
& $-2.25$ &86.2&
&$ -2.18$ &104.3\\

Eu {\tiny II}& 4205.04& 0.00 & $0.21$
& . . .& . . .&
& . . .& . . .&
&$<-2.25$ &$<$53.4\\
 \hline
  \end{longtable}
}

 \longtab[2]{ 
          \begin{longtable}{lccrrrp{0.05pt}rr}
           \caption{Line Data and Equivalent Widths of S~24-72 and S~11-04}
    \label{tab:linelistUVES}\\
            \hline
            \noalign{\smallskip}
& & & &\multicolumn{2}{c}{S$24-72$} & & \multicolumn{2}{c}{S$11-04$} \\
\cline{5-6}\cline{8-9}\\
Elem.& $\lambda$& L.E.P& log {\it gf}& log $\epsilon$&EW& &log $\epsilon$ & EW\\
&(\AA) &(eV) &  &  & (m\AA)& & &(m\AA)\\
\hline
\endfirsthead
\caption{continued.}\\
 \hline
            \noalign{\smallskip}
& & & &\multicolumn{2}{c}{S$24-72$} & & \multicolumn{2}{c}{S$11-04$} \\
\cline{5-6}\cline{8-9}\\
Elem.& $\lambda$& L.E.P& log {\it gf}& log $\epsilon$&EW& &log $\epsilon$ & EW\\
&(\AA) &(eV) &  &  & (m\AA)& & &(m\AA)\\
\hline
            \noalign{\smallskip}
\endhead
\hline
\endfoot
            \noalign{\smallskip}


Mg {\tiny I}&5172.68 & 2.71 & $-0.45$
& 4.79& 200.5&
& 5.00& 249.4\\

Mg {\tiny I}&5183.60& 2.72 & $-0.24$
& 4.91&237.4&
& 5.06& 293.8\\	

Mg {\tiny I}& 5528.40 &4.35 & $-0.50$
& 4.93 & 53.3&
& 5.03 & 67.8\\

Ca {\tiny I}&5588.76&2.53 & 0.36 
& 3.50& 45.1 &
& 3.51 & 51.7 \\

Ca {\tiny I}&5594.47&2.52 & 0.10 
&. . .& . . .&
& 3.71 &47.4 \\

Ca {\tiny I}&6102.72&1.88 & $-0.77$ 
&. . .& . . .&
& 3.65 &44.0 \\

Ca {\tiny I}&6122.22&1.89 & $-0.32$ 
& 3.73& 74.1&
&3.72& 84.5\\

Ca {\tiny I}&6162.17&1.90 & $-0.09$ 
& 3.39& 65.0&
& 3.79 & 108.3  \\

Ca {\tiny I}&6439.07&2.53 & $0.39$ 
& 3.53& 53.3&
&. . .&. . . \\

Sc {\tiny II}& 5031.02& 1.36 & $-0.40$
& $0.01$ & 40.4&
& $0.14$ & 54.1\\

Sc {\tiny II}& 5526.79& 1.77 & $0.02$
& 0.20&45.9&
& $0.19$& 50.1\\

Ti {\tiny I}& 4981.73& 0.85 &0.57
& 1.82& 54.6&
&1.91& 76.0\\

Ti {\tiny I}& 4991.07& 0.84 &0.45
& 1.67 & 37.8&
& 1.99& 74.4 \\

Ti {\tiny I}& 4999.50& 0.83 &0.32
& 1.92 & 46.3&
& 1.92& 59.6 \\

Ti {\tiny I}& 5007.21& 0.82 &$0.17$
& . . . & . . . &
 & 2.01 & 56.1 \\

Ti {\tiny I}& 5064.65& 0.05 &$-0.94$
& . . . & . . . &
 & 1.98& 49.6 \\

Ti {\tiny I}& 5173.74& 0.00 &$-1.06$
& . . . & . . . &
 & 2.12 & 57.1 \\

Ti {\tiny I}& 5192.97& 0.02 &$-0.95$
& 1.98& 37.0&
 &2.20& 70.8 \\

Ti {\tiny I}& 5210.38& 0.05 &$-0.82$
& . . . & . . . &
 & 2.20 & 78.9 \\

Ti {\tiny II}&4443.80 & 1.08 & $-0.71$
& 2.22 & 142.0&
& . . . & . . .  \\

Ti {\tiny II}&4798.53& 1.08 &$ -2.66$ 
& 2.36& 33.8&
& 2.35 & 38.8 \\

Ti {\tiny II}&4865.61& 1.12 &$ -2.70$ 
& 2.15& 20.1&
& . . . & . . . \\

Ti {\tiny II}& 5129.16& 1.88 &$-1.34$
& . . . & . . . &
 & 2.27 & 53.3 \\

Ti {\tiny II}& 5154.07& 1.57 &$-1.78$
& . . . & . . . &
 & 2.40 & 61.1 \\

Ti {\tiny II}&5185.90& 1.89& $-1.41$ 
& 2.37& 49.7&
& 2.43& 59.5 \\

Ti {\tiny II}& 5336.79& 1.58 & $-1.60$ 
& 2.33 & 63.3&
& 2.46& 80.4 \\

Cr {\tiny I}& 5206.04& 0.94 & $0.02$ 
& 2.52& 114.5&
& 2.56& 135.7\\	

Cr {\tiny I}& 5208.42& 0.94 & $0.17$ 
& . . .& . . .&
& 2.57& 147.0\\	

Cr {\tiny I}& 5345.80& 1.00 & $-0.95$ 
& 2.37& 33.5&
& 2.57 & 58.6\\	

Cr {\tiny I}& 5409.77& 1.03 & $-0.67$ 
& 2.33& 47.1&
& 2.56 & 78.3 \\	

Mn {\tiny I}& 4041.35& 2.11 & $0.28$ 
&2.12&49.0&
&. . . &. . . \\

Mn {\tiny I}& 4783.43& 2.30 & $0.04$ 
&2.22&30.4&
&2.03&27.4 \\

Mn {\tiny I}& 4823.52& 2.32 & $0.14$ 
&2.30&38.5&
&2.15&37.4 \\

Fe {\tiny I}& 4260.47 &2.40 & $0.08$ 
&4.39& 134.2& 
&. . . & . . . \\

Fe {\tiny I}& 4859.74& 2.88 & $-0.76$ 
& 4.39 &58.8&
&. . . & . . . \\

Fe {\tiny I}& 4871.32& 2.87 & $-0.36$ 
& 4.51 & 93.7 &
& 4.52& 106.4\\

Fe {\tiny I}& 4872.14 & 2.88 & $-0.57$ 
&4.34& 67.6& 
&4.55& 93.0\\

Fe {\tiny I}& 4890.76& 2.88 & $-0.39$ 
& 4.46& 88.1 &
& 4.70& 114.7 \\

Fe {\tiny I}& 4891.49& 2.85 & $-0.11$ 
& 4.22 & 92.8 &
& 4.58& 127.2 \\

Fe {\tiny I}& 4918.99& 2.87 & $-0.34$ 
& 4.27 & 80.2&
& 4.66 & 117.0 \\

Fe {\tiny I}& 4920.50 & 2.83 & $0.07$ 
&. . . & . . .& 
& 4.33& 124.9\\

Fe {\tiny I}& 4939.69& 0.86 & $-3.34$ 
& 4.96 &102.5&
&. . . & . . . \\

Fe {\tiny I}& 4966.09 & 3.33 & $-0.87$ 
&. . . & . . .& 
& 4.78& 47.2\\

Fe {\tiny I}& 4994.13& 0.92 & $-2.96$ 
& 4.52 & 94.4 &
& 4.61 & 122.5 \\

Fe {\tiny I}& 5006.12& 2.83 & $-0.61$ 
& 4.42 & 76.0 &
& 4.57 & 97.8 \\

Fe {\tiny I}& 5012.07& 0.86 & $-2.64$ 
&. . . & . . .& 
& 4.73& 160.6 \\

Fe {\tiny I}& 5041.07& 0.96 & $-3.09$ 
& 4.71 & 94.9 &
& 4.84 & 126.0 \\

Fe {\tiny I}& 5041.76 & 1.49 & $-2.20$ 
& 4.52 & 94.6 &
& 4.41& 105.2 \\

Fe {\tiny I}& 5049.82 & 2.28 & $-1.34$ 
&. . . & . . .& 
& 4.56& 98.8 \\

Fe {\tiny I}& 5051.64& 0.92 & $-2.80$ 
& 4.50 &104.7 &
& 4.64& 137.5 \\

Fe {\tiny I}& 5068.77 & 2.94 & $-1.04$ 
&. . . & . . .& 
& 4.49& 51.5 \\

Fe {\tiny I}& 5079.74 &0.99 & $-3.22$ 
&4.72& 84.6& 
& 4.81& 111.3\\

Fe {\tiny I}& 5083.34& 0.96 & $-2.96$ 
& 4.57 &95.2 &
& 4.56& 115.5 \\

Fe {\tiny I}&5098.70 & 2.18 & $-2.03$ 
&4.67& 54.6& 
& 4.69 & 68.7 \\

Fe {\tiny I}& 5123.72 & 1.01 & $-3.07$ 
&4.58 & 83.5& 
& 4.71& 113.9 \\

Fe {\tiny I}& 5127.36& 0.92 & $-3.31$ 
& 4.78& 89.9 &
& 4.75 & 109.3 \\

Fe {\tiny I}& 5142.93& 0.96 & $-3.08$ 
& 4.77 & 101.3 &
& . . . & . . . \\

Fe {\tiny I}& 5150.84& 0.99 & $-3.07$ 
& 4.57 & 85.3 &
& 4.66 & 112.7 \\

Fe {\tiny I}& 5151.91& 1.01 & $-3.31$ 
&. . . & . . .& 
& 4.71& 94.5 \\

Fe {\tiny I}& 5166.28& 0.00 & $-4.20$ 
&4.79& 116.3& 
&4.88 & 157.3\\

Fe {\tiny I}& 5171.60& 1.49 & $-1.79$ 
&4.71 & 134.8& 
& 4.54& 145.8 \\

Fe {\tiny I}& 5191.46& 3.04 & $-0.55$ 
&. . . & . . .& 
& 4.56 & 84.4 \\

Fe {\tiny I}& 5194.94& 3.23 & $-2.22$ 
&4.81& 116.5& 
& 4.80 & 136.4 \\

Fe {\tiny I}& 5202.34& 2.18 & $-1.84$ 
&4.75 & 74.9& 
& 4.95 & 103.6\\

Fe {\tiny I}& 5216.27 & 1.61 & $-2.15$ 
&4.77 & 106.0& 
& 4.94 & 137.7\\

Fe {\tiny I}&5225.53 & 0.11 & $-4.79$ 
&5.14& 88.3& 
& 4.82 & 88.6 \\

Fe {\tiny I}&5254.96 & 0.11 & $-4.76$ 
&5.04 & 83.3& 
& 4.85 & 94.1 \\

Fe {\tiny I}& 5266.56& 3.00 & $-0.39$ 
& 4.65& 94.5 &
& . . . & . . . \\

Fe {\tiny I}& 5324.18& 3.21 & $-0.10$ 
& 4.49& 85.0 &
& . . . & . . . \\

Fe {\tiny I}& 5339.93 & 3.27 & $-0.65$ 
&4.55& 46.3& 
& 4.70& 66.0\\

Fe {\tiny I}& 5341.02& 1.61 & $-1.95$ 
&4.64 & 112.7& 
& . . . & . . . \\

Fe {\tiny I}& 5446.92 &0.99 & $-1.91$ 
&4.18 & 141.5& 
& 4.44& 192.1 \\

Fe {\tiny I}& 5455.61& 1.01 & $-2.10$ 
&4.62& 156.5& 
& 4.69 & 195.0 \\

Fe {\tiny I}& 5497.52& 1.01 & $-2.85$ 
& 4.51 & 99.1&
& 4.47 & 118.7 \\

Fe {\tiny I}& 5501.47& 0.96 & $-3.05$ 
& 4.56& 93.3&
& 4.71& 128.5 \\

Fe {\tiny I}& 5506.78& 0.99 & $-2.80$ 
&4.59& 110.5& 
& 4.63 & 138.5 \\

Fe {\tiny I}& 5586.76& 3.37 & $-0.10$ 
&. . . & . . .& 
& 4.51 & 85.5 \\

Fe {\tiny I}& 5615.64& 3.33 & $0.05$ 
&4.31& 74.4& 
& 4.54 & 102.9 \\

Fe {\tiny I}& 6136.61& 2.45 & $-1.40$ 
&4.56& 74.7& 
& 4.60 & 92.7 \\

Fe {\tiny I}& 6137.69& 2.59 & $-1.40$ 
&4.67& 69.1& 
& 4.63 & 79.4 \\

Fe {\tiny I}& 6230.72& 2.56 & $-1.28$ 
&4.77 & 89.6& 
& 4.64 & 95.0 \\

Fe {\tiny I}& 6252.56& 2.40 & $-1.77$ 
&. . . & . . .& 
& 4.78 & 84.2 \\

Fe {\tiny I}& 6393.60& 2.43 & $-1.43$ 
&4.59 & 78.1& 
&. . . & . . . \\

Fe {\tiny I}& 6421.35& 2.28 & $-2.03$ 
&4.81 & 64.5& 
&. . . & . . . \\

Fe {\tiny I}& 6430.85& 2.18 & $-2.01$ 
&4.75& 72.3& 
& 4.98 & 109.3 \\

Fe {\tiny I}& 6494.98& 2.40 & $-1.24$ 
&4.37& 80.4& 
& 4.46 & 104.8 \\

Fe {\tiny I}& 6592.91& 2.73 & $-1.47$ 
&4.53& 42.1& 
&. . . & . . . \\

Fe {\tiny II}& 4178.86 & 2.58 & $-2.51$ 
&4.29& 45.7& 
& . . . & . . . \\

Fe {\tiny II}& 4923.93& 2.89 & $-1.26$ 
&4.46&115.2& 
& 5.18 & 117.6\\

Fe {\tiny II}& 5018.45& 2.89 & $-1.10$ 
&. . . & . . .& 
& 5.16& 125.8 \\

Fe {\tiny II}& 5276.00& 3.20 & $-2.01$ 
&4.60& 53.0& 
& . . . & . . . \\

Fe {\tiny II}& 6247.55& 3.89 & $-2.51$ 
&. . . & . . .& 
& 5.17& 15.1 \\
		
Ni {\tiny I}& 5476.90 & 1.83 &$-0.78$
&3.35& 99.6 &
& 3.12& 100.4\\

Zn {\tiny I}& 4810.53& 4.08 & $-0.17$
&2.09&30.9&
& 2.09 &33.1 \\

Sr {\tiny II}&4077.71& 0.00 & 0.17 
& $-1.03$ &151.5 &
& $-0.63$ & 218.3\\

Sr {\tiny II}&4215.52& 0.00 & $-0.17$
& . . . & . . . &
& $-0.60$& 185.6 \\

Ba {\tiny II}& 4934.10& 0.00 & $-0.16$
& $-2.17$& 85.7&
& . . . & . . .  \\

Ba {\tiny II}& 6141.73& 0.70 & $-0.08$
& $-2.00$ & 35.0 &
& $-1.66$& 67.5 \\

Ba {\tiny II}& 6496.91& 0.60 & $-0.38$
& $-1.95$ & 28.5&
& $-1.68$ & 53.3\\

Eu {\tiny II}& 4205.04& 0.00 & $0.21$
&$<-2.60$ &$<$34.7&
&$<-2.67$ &$<$36.8\\

 \hline
  \end{longtable}
}
\end{appendix}

\end{document}